\documentclass[epjST]{svjour}

\pdfoutput=1
\usepackage{amsmath,amssymb,xspace,graphicx,xcolor}
\usepackage{hyperref}
\usepackage{ulem}
\renewcommand{\r}{{\bf r}}
\newcommand{\n}{{\nabla}}
\newcommand{\p}{{\bf p}}

\renewcommand{\u}{{\bf u}}

\newcommand{\J}{{\bf J}}
\newcommand{\Q}{{\bf Q}}
\newcommand{\V}{{\bf V}}

\newcommand\half{\frac{1}{2}}

\newcommand\calf{{\cal F}}

\DeclareMathOperator{\Tr}{Tr}
\newcommand{\ra}{\rangle}
\newcommand{\la}{\langle}

\newcommand\dO{{\rm d}\Omega}

\newcommand\grad{\nabla}
\renewcommand\d{{\rm d}}
\newcommand\A{{\bf A}}

\begin{document}

\title{Active Brownian Particles and Run-and-Tumble Particles: a Comparative Study}
\author{A. P. Solon\inst{1}, M. E. Cates\inst{2}, J. Tailleur\inst{1}}
\institute{\inst{1}~Univ Paris Diderot, Sorbonne Paris {Cit\'e}, MSC, UMR 7057 CNRS, F75205 Paris, France\\ \inst{2}~SUPA, School of Physics and Astronomy, University of Edinburgh, Edinburgh EH9 3FD, United Kingdom }

\date{\today}
 
\abstract{Active Brownian particles (ABPs) and Run-and-Tumble
  particles (RTPs) both self-propel at fixed speed $v$ along a
  body-axis ${\bf u}$ that reorients either through slow angular
  diffusion (ABPs) or sudden complete randomisation (RTPs). We compare
  the physics of these two model systems both at microscopic and
  macroscopic scales. Using exact results for their steady-state
  distribution in the presence of external potentials, we show that
  they both admit the same effective equilibrium regime perturbatively
  that breaks down for stronger external potentials, in a
  model-dependent way. In the presence of collisional repulsions such
  particles slow down at high density: their propulsive effort is
  unchanged, but their average speed along ${\bf u}$ becomes $v(\rho)
  < v$. A fruitful avenue is then to construct a mean-field
  description in which particles are ghost-like and have no
  collisions, but swim at a variable speed $v$ that is an explicit
  function or functional of the density $\rho$. We give numerical evidence that the
  recently shown equivalence of the fluctuating hydrodynamics of ABPs and RTPs
  in this case, which we detail here, extends to microscopic
  models of ABPs and RTPs interacting with repulsive forces.}

\maketitle

\section{Introduction}

Outside the realm of exact results, and despite recent
progress~\cite{Derrida2007}, non-equilibrium statistical mechanics
largely remains (paraphrasing Harish-Chandra~\cite{Kac}) messy,
elusive and non-rigorous. However, we may say (paraphrasing Dyson~\cite{Kac}) that this is
precisely why it is such an interesting research field. While a fully general extension
of the well-established framework of equilibrium
statistical physics remains out of reach, there are important classes of
non-equilibrium systems for which progress towards a comprehensive theory
seems achievable. Colloidal self-propelled particles (SPPs), which represent
a central focus of research into active matter
\cite{sriram,RoP,romanczuk,RMP},  arguably offer such a prospect. This applies at least in some
simplified limits, such as spherical SPPs interacting by central forces only (a restriction that
excludes hydrodynamic interactions). In developing theories of such systems, an important general question 
is how far macroscopic behaviour is sensitive to the detailed dynamical rules.
For thermal systems, steady-state properties are governed by Boltzmann equilibrium, and hence sensitive to the energy landscape but
not the dynamics used to explore it -- so long as this dynamics obeys the principle of detailed balance (as it must do in such cases). 
For non-equilibrium systems, such as SPPs, the question must be asked afresh. In this article we
address this issue, by
studying two classes of active particles, ABPs and RTPs. These have related but distinct dynamics, and have
become two major workhorses of recent simulation and theoretical studies of active matter. We will 
find instances where their dynamical differences disappear on coarse-graining, to give the same macroscopic physics; and other instances where these differences remain important -- typically when there is short-scale structure in the problem, such as confinement in small traps.

In active matter, a continuous supply of energy destroys microscopic
time-reversal symmetry (TRS) and allows phenomena to arise that are impossible
in thermal equilibrium systems, where detailed balance restores
TRS in the steady state. Inspired in large part by
experiments on synthetic self-propelled colloidal particles~\cite{PennState,Jones,golestanian,Palacci2010PRL,Bocquet2012,bechinger,palacci,herminghaus},
many recent theory and simulation papers have addressed the physics of
a simplified model comprising spherical active Brownian particles
(ABPs)~\cite{fily,baskaran,sten1,gompper,sten2,speck,redner,mallory,Cristina,Brady,Henkes}. In
parallel, growing experimental interest in bacterial
motion~\cite{Galajda2007,Leonardo2010PNAS,Sokolov2010PNAS,Saragosti2011PNAS,Hwa1,Vladescu2015PRL,SchwarzLinek2012PNAS}
has brought run-and-tumble particles (RTPs) to the fore as a second prototypical model of
self-propelled particles at the colloidal scale.

In these models, each particle feels a constant external force $\zeta
v_0$ oriented along its swim direction ${\bf u}$. At low density this
is equal and opposite to the local drag force felt by a particle
moving at speed $v_0$. (At high density the external force may be
partly opposed instead by conservative interparticle forces.) This is
a one-body drag and thus the models entirely omit hydrodynamic
interactions and all the other physics associated with the presence of
an incompressible solvent. These can have important effects, for
instance generating self-pumping states~\cite{Nash2010,Stark1} or
causing rotation by collision~\cite{Fielding,Stark}. On the contrary,
in the spherical ABP model the angular dynamics comprises rotational
diffusion with a fixed diffusivity $D_r$.  Idealized RTPs differ from
ABPs only in their rotational relaxation; instead of continuous
diffusion, RTPs undergo discrete tumbling events of short duration
that completely randomize their swimming direction; these events occur
randomly in time at mean rate $\alpha$ and the mean time between
  two tumbles is therefore exponentially distributed~\footnote{Other
    type of distributions, possibly with large tails, may also be
    considered. See,
    e.g.,~\cite{Cluzel2004Nature,ThielPRE2012}}. Both models share an
important simplification, which would be inadmissible for any
particles capable of exerting significant torques on one another (such
as non-spherical SPPs at high density). Specifically, the angular
dynamics of each particle remains unperturbed by interactions of any
kind, and proceeds independent of the positions or orientations of all
other particles in the system.  In what follows, the terms ABP and RTP
refer to this spherical, torque-free case unless stated otherwise.

Another key feature of these two models is that each particle
exchanges momentum with external driving and drag forces, not with a
suspending solvent. The resulting non-conservation of momentum within
the system greatly complicates the discussion of the macroscopic force
balances that normally underly mechanical properties such as pressure
\cite{mallory,Cristina,Brady}. Indeed, the force per unit area on a
wall in general depends on the wall-particle
interactions~\cite{Solon1}; this dependence does cancel out for the
case of spherical (torque-free) ABPs moving with constant propulsive
force~\cite{Solon2}, but not if the propulsive force is itself
density-dependent~\cite{Solon1} (a case we return to below).

In this paper, we present a comparative study of these simplified
models of self-propelled particles. We first consider the steady-state
distribution of non-interacting SPPs submitted to an external
potential $V_{\rm ext}$. Using exact results on the sedimentation of
active particles, we show that ABPs and RTPs each admit an ``effective
temperature" regime when the Stokes speed $v_s\equiv -\nabla V_{\rm
  ext}/\zeta$ is small compared to the swim speed $v$. We show how the effective
temperature concept breaks down outside this regime in model-specific ways; this issue
has attracted lots of interest
recently~\cite{Palacci2010PRL,Stark1,TailleurCates2009EPL,Szamel2014PRE,Ginot2011PRX}. We
then consider the confinement of SPPs in circular or spherical traps, and show how the different angular
dynamics between the two systems result in qualitatively different
behaviours outside the effective equilibrium regime.

We then turn to interacting SPPs and consider models that replace
interparticle collisions with a density-dependent propulsion
force. Therefore we avoid any detailed discussion of the mechanical
pressure, although it may offer an interesting alternative perspective
on phase equilibria \cite{Brady,Solon2,Brady2}. We explore an approximation scheme 
for torque-free SPPs with collisional
interactions whereby the conservative forces responsible for
collisional slow-down of the particles are replaced at mean-field
level by a ``programmed'' slow-down, that is, an effective reduction
in the propulsive force at high density. This path was first sketched
out in \cite{TC} and pursued further in
\cite{sten1,sten2,CT,Thompson2011JSM,ModelB}. In a more general context
it is both legitimate and interesting to consider, in its own right,
the case with no conservative force field between particles, and ask
about the effects of programmed slow-down among such ``ghost''
particles. While not forgetting its original motivation in the
collisional case, we will often adopt this viewpoint here.
Indeed, the concept of a programmed slow-down is well established in
biological systems such as bacteria, where density-dependent dynamics
can be effectively introduced through a biochemical pathway called
quorum sensing~\cite{RoP,Hwa1,Hwa2,quorum}. This causes bacteria to
change behaviour in response to the concentration of a short-lived
chemical that they also emit. Because of the short lifetime, the
response is then directly sensitive to the number of neighbours within
a short but finite distance, and hence to a local coarse-grained
density. Note that this is distinct from
chemotaxis~\cite{Berg,Schnitzer} in which organisms swim
preferentially up or down \textit{gradients} of a long-lived chemical
signal---effectively creating a long-range repulsion or
attraction. (Chemotaxis might also arise in synthetic, self-phoretic colloids
\cite{SriramGolestanian}.)  An interesting study of chemotactic
interactions in ABPs, showing some features resembling phase
separation of the type discussed below, can be found in \cite{Meyer}.

In bacteria such as \textit{E. coli} angular relaxation is an RTP
process, for which the consequences of density-dependent speed were
first worked out in \cite{TC}. However, some bacterial types,
including mutant strains of \textit{E. coli} known as ``smooth
swimmers", do not tumble. In practice the genetic engineering
strategies used in~\cite{Hwa1,Hwa2} to give density-dependent
slow-down primarily achieve this by altering tumble times and
frequencies (reducing the time-averaged speed). However, were it
possible to directly link the actual propulsion speed of bacteria to
their local density in a smooth-swimming strain, then this would offer
an unambiguous physical realization of ABPs (as opposed to RTPs) with
a density dependent speed. As in~\cite{Hwa1,Hwa2}, this could probably
be made to happen at densities too low for collisional or crowding
effects to be important.

In summary, the study of SPPs with density-dependent propulsion speed
is motivated (i) as an approximate representation of the collisional
slow-down in SPPs with pairwise repulsions; (ii) as a representation
of smooth-swimming or run-and-tumble bacteria with slow-down caused by
quorum sensing or a related, non-collisional mechanism; and (iii) as a
model in its own right, with which to explore generic collective
phenomena in active matter systems. In the last context, it offers further insights
into the degree to which macroscopic outcomes depend on local dynamical rules, 
and we shall focus on this below in comparing the ABP and RTP cases
in some detail.  

The most striking many-body phenomenon seen in this class of models is
motility-induced phase separation, or MIPS. This is by now a
well-established concept
\cite{RoP,TC,CT,Thompson2011JSM,ModelB,Farrell2012PRL,MIPS,PNAS,Loewen1,Loewen2a}
and is the result of two effects in combination. First, the mean speed
of a motile particle along its propulsion direction decreases with
density (in contrast to the equilibrium case where the velocity
statistics of a particle are separable, and fixed solely by equipartition). Second, particles tend to accumulate in
spatial regions where they move slowly (a possibility ruled out, in
isothermal equilibrium, by the fact that the speed distribution cannot
depend on position). This combination creates a positive feedback
which is the origin of MIPS \cite{RoP,TC,MIPS}. 

In what follows, after exploring in Section \ref{extpot}
the one-body behaviour of ABPs and RTPs in external potentials, we turn in Section \ref{mbp}
to their many-body physics, showing first how
the large-scale dynamics  with density-dependent
$v(\rho)$ can in each case be mapped on the equilibrium dynamics of passive
particles with attractive forces, and then outlining how gradient terms affect
this mapping. Sections \ref{extpot} and \ref{mbp} offer a more complete presentation
of results reported in \cite{TailleurCates2009EPL,CT} respectively. In fact we
add substantially to these results: our discussion of ABPs in traps and of angular
distributions for ABPs and RTPs under gravity are both new to the current paper.
In the many-body section we carefully compare our approach with recent work
by others on the role of leading order nonlocality (gradient corrections) in an effective free energy picture. We
show further that, somewhat surprisingly,
the large-scale equivalence between ABPs and RTPs established at the
level of fluctuating hydrodynamics for particles with density-dependent
  swim speed~\cite{CT} effectively extends to \textit{microscopic simulations} of
ABPs and RTPs interacting with \textit{repulsive forces}.


\section{Steady-state of active particles in external potential}

\label{extpot}

Consider the classical experimental geometry, devised by Jean Perrin,
in which dilute colloidal particles are allowed to sediment under
gravity in a cuvette. Equilibrium statistical mechanics asserts that
we can forget all details of the diffusive dynamics of the colloids:
the steady-state probability of finding a particle at a given height
$z$ is a Boltzmann distribution, $P(z)\propto \exp(-\delta m g z/kT)$,
where $\delta m$ is the mass difference between the particle and a
corresponding volume of liquid~\footnote{Note that in all this article
  but the sedimentation sections, we silently assume the mass of
  particles to be equal to one.}. Let us now replace the Brownian
random-walk diffusion of the colloids by an active random walk,
powered by some microscopic non-equilibrium process. In the SPP
context the latter is a persistent random walk of fixed speed, with
either continuous (ABP) or discrete (RTP) changes of direction. (Its
persistence length is much longer than for true Brownian motion,
although the latter is finite in principle. More importantly, the
instantaneous speed of true Brownian motion is not fixed, but has
unbounded fluctuations given by the Maxwell-Boltzmann distribution.)
Though not identical, clearly the equilibrium and non-equilibrium
diffusive processes cannot be completely unrelated, at least in a
limit where the persistence length is much smaller than the
sedimentation length. Outside this regime, however, very different
physics may be encountered.

Previous studies of sedimenting active particles confirm this line of
reasoning. Exact results for RTPs in one, two and three spatial
dimensions have shown the density profile to be exponential
$\rho(z)\propto \exp(-\lambda z)$ far away from the containing
boundaries~\cite{TailleurCates2009EPL,TC}. The sedimentation length
reduces to $\lambda^{-1}=k T_{\rm eff}/\delta m g$, with $T_{\rm
  eff}=v^2\zeta/ d \alpha$ and $\zeta$ the inverse mobility of the particle; this result holds in
the limit where the sedimentation speed $v_s=\zeta^{-1} \delta m g$ is much
smaller than the swim speed $v$. 

This effective equilibrium regime,
where the sedimentation profile is given by a Boltzmann weight with a
temperature $T_{\rm eff}$ independent of the potential $V_{\rm ext}$,
was observed experimentally for ABPs
in~\cite{Palacci2010PRL,Ginot2011PRX} and derived analytically
in~\cite{Stark1}. Its breakdown for
stronger confinement, predicted in~\cite{TailleurCates2009EPL}, was
not observed experimentally so far~\cite{Palacci2010PRL,Ginot2011PRX},
triggering a debate about the generic sedimentation behaviour 
for SPPs~\cite{Szamel2014PRE}. Below, for both RTPs and ABPs (Sections \ref{2DR},\ref{2DA}) we derive in 2D 
the full probability distribution $P(z,\theta)$ for finding a particle at 
height $z$ swimming in direction $\theta$. The outcome is that
ABPs and RTPs both admit the same effective equilibrium regime -- up to a simple
parameter mapping between $\alpha$ and $D_r$. However, beyond this regime the
density profiles are dynamics-dependent, and various definitions of the
``effective temperature'', which all coincide within the regime of small $v_s$, are found to differ beyond it.

In Section \ref{TRAP} we turn from sedimentation (which has $v_s$ uniform in space) 
to address the steady-state distributions of ABPs and RTPs in
harmonic traps, where $v_s(r)=-\kappa r$. Again, ABPs and RTPs yield
the same limiting effective equilibrium regime but lead to very different
physics once the persistence length (respectively $v/\alpha$ or $v/D_r$) is much
larger than the trap radius $v/\kappa$. In this regime, the cases of continuous (ABP) and discrete (RTP) angular rotation
give entirely different statistics for the density at the centre of the trap, for reasons we discuss.

\subsection{Sedimentation of 2D RTPs}\label{2DR}
Let us consider the dynamics of a 2D RTP in the presence of an
external potential $V_{\rm ext}({\bf r}) = \delta m g z$. The self-propulsion
velocity $v{\bf u}(\theta)$ is supplemented by a sedimentation velocity
$-v_s {\bf e}_z=-\zeta^{-1}\nabla V_{\rm ext}({\bf r})$ so that the master
equation reads
\begin{equation}
  \dot P(\r,\theta)=-\nabla \cdot [(v {\bf u}(\theta)-v_s {\bf e}_z) P(\r,\theta)]-\alpha P(\r,\theta)+\frac{\alpha}{2\pi} \int d\theta' P(r,\theta')
\label{ME} \end{equation}
We consider a system which is not free-falling, i.e. such that a wall
at $z=0$ prevents the particles from crossing this plane. 
A number of boundary conditions can be
used to model the relevant interaction; for example,
one can simply cancel the $z$ component of the particles
velocity~\cite{Wan2008PRL}, or take into account the torque that a
wall would exert on actual bacteria~\cite{TailleurCates2009EPL}. Different boundary layer structures
then arise proximal to the wall, depending on the details of the dynamics and the boundary condition, which typically
show strong particle accumulation in this region~\cite{Elgeti1,Elgeti2}. Beyond 
this proximal regime (typically a few run lengths in height)  a stationary profile is
reached at larger $z$ whose form is independent of the boundary condition. 

\begin{figure}
  \begin{center}
    \includegraphics{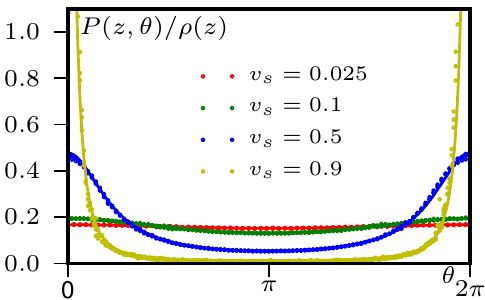}
    \includegraphics{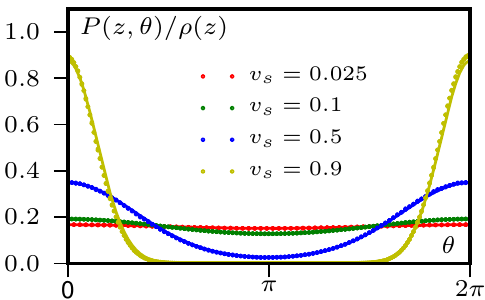}\\[5mm]
    \includegraphics{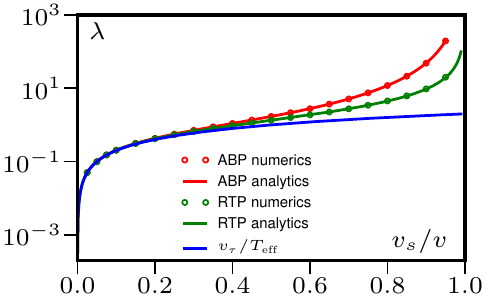}
    \includegraphics{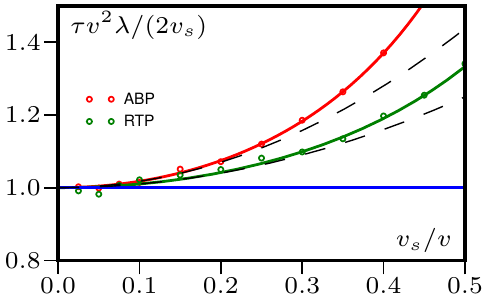}
  \end{center}
  \caption{Sedimentation of RTPs and ABPs ($\alpha=D_r=v=1$). Symbols
    correspond to simulations data and full lines to analytical
    predictions. The conditional probability $P(\theta|z)$ for ABPs
    ({top left}) and RTPs ({top right}) is independent of $z$ in the
    distal part of the profile. (For each value of $v_s$, several
    $P(\theta|z)$ are constructed for different $z$ in the distal
    region; they show perfect overlap.) The simulation data matches
    the theoretical predictions given by~\eqref{eqn:fthetaRTP}
    and~\eqref{eqn:fthetaABP}. {Bottom left}: Inverse sedimentation
    length $\lambda$ as $v_s/v$ is varied. The equivalence between
    ABPs and RTPs breaks down as soon as one leaves the effective
    equilibrium regime (bottom right). The dashed lines correspond to
    the first order corrections of the two systems given
    in~\eqref{lambdartp} and~\eqref{lambdaabp}.}
  \label{fig:sedim}
\end{figure}

In what follows we consider only this distal part
of the profile. Its form can be found by assuming a factorized
steady-state $P(z,\theta)=\rho(z) f(\theta)$; from Eq.~\ref{ME}, the factors must then satisfy
\begin{equation}
  \label{eqn:sepvarrtp}
  \frac{\rho'(z)}{\rho(z)}=\frac{\alpha}{v\cos\theta-v_s}\left(\frac{1}{2\pi f}-1\right)\equiv-\lambda
\end{equation}
where $\lambda$ is a constant, a priori
unknown. Eqs.~\eqref{eqn:sepvarrtp} directly give
\begin{equation}
  \rho(z)=\rho_0 e^{-\lambda z};\qquad f(\theta)=\frac{1}{2\pi \left[1-\frac\lambda\alpha (v\cos\theta-v_s)\right]}
\end{equation}
The constant $\lambda$ can then be fixed by the normalization condition
$\int_0^{2\pi}f(\theta)d\theta=1$, which yields
\begin{equation}
  \alpha \sqrt{\frac{\lambda(v_s+v)+\alpha}{\lambda(v_s-v)+\alpha}}=\lambda(v_s+v)+\alpha
\end{equation}
Finally, the inverse sedimentation length is given by
\begin{equation}
  \label{eqn:lambdaRTP}
  \lambda=\frac{2\alpha v_s}{v^2-v_s^2}
\end{equation}
so that
\begin{equation}
  \label{eqn:fthetaRTP}
  f(\theta)=\frac{1-\frac{v^2_s}{v^2}}{2\pi \left[1-2 \frac{v_s}v \cos\theta+\frac{v_s^2}{v^2})\right]}
\end{equation}
These exact results extend those of~\cite{TailleurCates2009EPL} where only
$\rho(z)$ was computed; they perfectly fit simulations of
sedimenting RTPs, as shown in figure~\ref{fig:sedim}.

\subsection{Sedimentation of 2D ABPs}\label{2DA}
A similar path can be followed for sedimenting ABPs in 2D, starting
from the master equation
\begin{equation}
\dot P(\r,\theta)=-\nabla \cdot [(v {\bf u}(\theta)-v_s {\bf e}_z) P(\r,\theta)] +D_r \partial_{\theta\theta} P(r,\theta)
\label{ME2}\end{equation}
Again, a separation of spatial and angular variables is confirmed numerically beyond a boundary
layer proximal to the wall. The steady-state $P(z,\theta)$ is found by
using the ansatz $P(z,\theta)=\rho(z)f(\theta)$ in Eq.\ref{ME2}, to give
\begin{equation}
  0=-\partial_z[(v\cos\theta-v_s) P(z,\theta)]+D_r\partial_{\theta\theta}P(z,\theta)
\end{equation}
which yields
\begin{equation}
  \rho'(z)=-\lambda \rho(z)\qquad f''(\theta)=-\frac{\lambda}{D_r}(v\cos\theta-v_s)f(\theta)
\end{equation}
As before, this yields an exponential atmosphere
$\rho(z)=\rho_0\exp(-\lambda z)$ where the inverse sedimentation length
$\lambda$ is set by the normalisation of $f(\theta)$. The equation for
$f(\theta)$ is the Mathieu equation. Given that $f(\theta)$ is even, it admits the general solution
\begin{equation}
  \label{eqn:fthetaABP}
  f(\theta)=c_1 C\left(-\frac{4\lambda v_s}{D_r},-\frac{2\lambda v}{D_r},2\theta\right)
\end{equation}
where $C(a,q,{\vartheta})$ is the cosine Mathieu function~\cite{dlmf}. For fixed $q$, this is
$\pi$-periodic in ${\vartheta}$ only for a countable subset of $a$ values, $a=a_n(q)$, which satisfy
$a_n(0)=n^2$. Since $\lambda\to 0$ when $v_s\to 0$, the solution of interest has
$n=0$ which implies
\begin{equation}
-4 \frac{\lambda v_s}{D_r}=  a_0\left(-2\frac{\lambda v}{D_r}\right)
\end{equation}
Solving this numerically gives $\lambda(v_s, v,D_r)$. (The numerics can be done using either a
  formal solver like Mathematica or by procedures based on
  series expansion of the $a_n$'s~\cite{refsmathieu}.)  The resulting values
of $\lambda$ match perfectly the results found by simulating sedimenting ABPs; the full distribution is then
obtained by plotting the corresponding Mathieu function, and also
agrees very well with the numerics (see Fig~\ref{fig:sedim}).

\subsection{Effective equilibrium regimes}\label{EER}
\paragraph{Boltzmann distributions.}
When $v_s/v\ll 1$, both ABPs and RTPs admit an effective
equilibrium regime. This is most easily seen for RTPs where $\lambda$ can be readily expanded:
\begin{equation}
\label{lambdartp}
  \lambda \underset{v_s\ll v}\simeq \frac{2\alpha v_s}{v^2} (1+\frac{v_s^2}{v^2})
\end{equation}
Keeping only the dominant contribution, the steady-state thus reduces to the Boltzmann form
\begin{equation}
  \rho(z)=\rho_0 e^{-V_{\rm ext}(z)/k T_{\rm eff}};\qquad k T_{\rm eff}=\frac{v^2\zeta}{2\alpha}
\end{equation}
where the last equality is an effective Stokes-Einstein relation that
connects the mobility $\zeta^{-1}$ to the diffusivity
$D_0={v^2}/{(2\alpha)}$ through an effective temperature $T_{\rm
  eff}$.

For ABPs, one can use the expansion
\begin{equation}
  a_0(q)\underset{q\ll 1}\simeq-\frac 1 2 q^2 + \frac 7 {128} q^4
\end{equation}
to get
\begin{equation}
  \label{lambdaabp}  \lambda \underset{v_s\ll v}\simeq \frac{2D_r v_s}{v^2} (1+\frac{7v_s^2}{4v^2})
\end{equation}
Again, the system admits a Boltzmann distribution as steady-state, at
the dominant order in $v_s/v$, with an effective temperature
\begin{equation}
  k T_{\rm eff}=\frac{v^2\zeta}{2D_r}
\end{equation}
Equating the mean time $\tau=\alpha^{-1}$ between two tumbles in RTPs
and the rotational diffusion time $\tau=D_r^{-1}$ in ABPs, the two
effective equilibrium regimes match. (In general dimensions, the
parameter mapping is from $\alpha^{-1}$ to $(d-1)D_r^{-1}$, which are
the angular autocorrelation times for RTPs and ATPs respectively
\cite{CT}.) However, even the first order corrections
in~\eqref{lambdartp} and~\eqref{lambdaabp} are different in form (see
Fig.~\ref{fig:sedim}). Matching the angular relaxation times gives
exponential decay of the angular correlator $\langle {\bf u}(t){\bf
  u}(0)\rangle = \exp[-t/\tau]$ in both cases, but unless $v_s/v \ll1$
the sedimentation length depends on the details of the angular
dynamics.

Note that one can also expand the angular distribution $f(\theta)$
close to the effective temperature regime $v_s\ll v$. For 2d RTPs,
this gives (using the normalisation $\int\d \theta  f(\theta)=1$)
\begin{align}
  2\pi f(\theta)&=1+\left(\frac{\kappa v}{\alpha}+\frac{3\kappa^2v v_s}{\alpha^2}\right)\cos\theta+\frac{\kappa^2 v^2}{2\alpha^2}\cos 2\theta+O(\kappa^2)\\
  &=1+\frac{2 v_s}{v}\cos \theta+\frac{2v_s^2}{v^2}\cos 2\theta+O(\left(\frac{v_s}{v}\right)^3)
\end{align}
where on the last line we used that
\begin{equation}
 \kappa \underset{v_s\ll v}\simeq \frac{2\alpha v_s}{v^2} (1+\frac{v_s^2}{v^2})
\end{equation}

We can do the same for 2d ABPs using the Fourier expansion of
characteristic Mathieu functions~\cite{dlmf}, to get
\begin{align}
  2\pi f(\theta)&=1+\frac{\kappa v}{D_r}\cos\theta+\frac{\kappa^2 v^2}{8D_r^2}\cos 2\theta+O(\kappa^2)\\
  &=1+\frac{2 v_s}{v}\cos \theta+\frac{v_s^2}{2v^2}\cos 2\theta+O(\left(\frac{v_s}{v}\right)^3)
\end{align}
where on the last line we used that
\begin{equation}
 \kappa \underset{v_s\ll v}\simeq \frac{2D_r v_s}{v^2} (1+\frac{7v_s^2}{4v^2})
\end{equation}
The difference between $f(\theta)$ for ABPs and RTPs can thus be
seen at the level of the second harmonics. As for the difference
between the density profiles of RTPs and ABPs, it appears at second
order in $v_s/v$. Note that the first order correction to
$f(\theta)=1/(2\pi)$ however suffices to distinguish these
distributions from equilibrium ones.

\paragraph{Fluctuation-dissipation relations.}
A quantity that is frequently looked at in non-equilibrium statistical
physics is the ratio between correlation and response
functions~\cite{Temperature}. Here we examine this for active
particles in a sedimentation setup. For simplicity we only consider
RTPs where we have explicit formulae, but conceptually what follows
applies equally to ABPs.  Let us consider a small force $f$ applied
along ${\bf e}_z$ and compute the response function
\begin{equation}
  R=\left.\frac{\partial \langle z \rangle}{\partial f}\right|_{f=0}
\end{equation}
The steady-state distribution is then $\rho(z)\propto
\lambda(f)\exp[-\lambda(f)z]$ where $\lambda(f)$ is obtained by
replacing $v_s\to v_s-\zeta^{-1} f$ in Eq.~\eqref{eqn:lambdaRTP}. Since $\la z \ra=\lambda(f)^{-1}$, $R$
is given by
\begin{equation}
  R=-\frac{\lambda'(0)}{\lambda(0)^2}
\end{equation}
Since the correlation function $C\equiv \langle z^2
\rangle-\langle z \rangle^2=\lambda(0)^{-2}$,  one directly gets
\begin{equation}
  \frac{C}{R}= -\frac{1}{\lambda'(0)}
\end{equation}
In the effective equilibrium regime, where $\lambda(f) = 2 (v_s-f/\zeta)/\tau v^2$
this gives
\begin{equation}\label{EFF}
  \frac{C}{R}= \frac{\zeta v^2\tau }{2}= kT_{\rm eff}
\end{equation}
which is the usual form for a system with an effective temperature;
the system satisfies a fluctuation dissipation theorem (as it should)
governed by that temperature.

For larger $v_s$, the same ratio can be computed explicitly for RTPs
using the exact result~\ref{eqn:lambdaRTP}, yielding
\begin{equation}
  \frac{C}{R}=\zeta\tau \frac{v^2-v_s^2}{2} \frac{v^2-v_s^2}{v^2+v_s^2}
\end{equation}
In this regime, there is no longer an FDT at temperature  $T_{\rm eff}$ for the response to a force along $z$. Moreover, an alternative definition of the effective temperature (which coincides with the fluctuation-dissipation ratio in the effective equilibrium regime) is found by applying a 
force instead along ${\bf e}_x$, and asserting a Stokes-Einstein relation between the mobility
$\zeta^{-1}$ and the lateral diffusivity defined via the mean-square displacement along $x$. This yields $kT_{\rm eff} = \zeta v^2\tau/2$, just as in Eq.\eqref{EFF}, with no dependence at all on $v_s$. For large $v_s$, the resulting ``horizontal" effective temperature controls neither the density profile 
nor the fluctuation-dissipation in the $z$ direction. 

In summary, outside the effective equilibrium regime, the three equally plausible definitions
\begin{equation}
  kT_{\rm eff}\equiv \frac{C}{R}\quad\text{and}\quad kT_{\rm eff}=-\frac{V_{\rm ext}(z)}{\log P(z)}\quad\text{and}\quad kT_{\rm eff}\equiv \zeta\lim_{t\to\infty}\langle x^2(t)/t \rangle
\end{equation}
yield three different effective temperatures. It follows that none is
really effective, in the sense that if a new one is needed to
describe each property measured, the concept of effective temperature is itself ineffective. In contrast, for both ABPs and RTPs within the effective
equilibrium regime (arising in the limit of weak gravity) all three
definitions of temperature coincide and the dynamics effectively
reduces to an equilibrium problem. For this reason we prefer the term
``effective equilibrium'' to describe this regime over the less
explicit ``effective temperature'' name used
earlier~\cite{TailleurCates2009EPL}.

Interestingly it was recently shown experimentally, by measuring
density profiles, that the addition of repulsive interactions between
the particles does not immediately destroy the effective equilibrium
regime~\cite{Ginot2011PRX}. Note that observing its breakdown
experimentally requires a rather large ratio of $v_s/v$ (for
$v_s/v\simeq 0.4$, the difference in $\lambda$ is only $40\%$ for
ABPs). This may explain why this breakdown has not been reported
experimentally so far. The more general question of effective
equilibrium in active particles remains a subtle one. For instance, if
the fixed self-propulsion speed $v$ is replaced by a fluctuating one,
using an Ornstein-Uhlenbeck process~\cite{Szamel2014PRE,Ginot2011PRX},
the effective equilibrium regime extends to arbitrary sedimentation
speed $v_s$, which clearly differs from the RTP and ABP cases treated
here.  As discussed in \cite{TailleurCates2009EPL}, fixed propulsion
speed is qualitatively different from any dynamics that can sample
very high speeds, even if these are rare (as they are in the thermal
Maxwell-Boltzmann distribution). One reason for this is clear: if
there is a fixed maximum speed, then when $v_s$ exceeds this, all
particles are moving only downwards and the sedimentation length must
be strictly zero (in the absence of particle-particle interactions).
Rare sampling of high molecular speeds is why, in thermal equilibrium,
a nonzero sedimentation length $\lambda^{-1} = kT/\delta m g$ is
maintained even for very large $v_s = \delta m g /\zeta$.

\subsection{Trapping of SPPs}
\label{TRAP}
Apart from RTPs in one dimension~\cite{TC}, there are no exact results
available for the full steady-state distribution of ABPs and RTPs in a
harmonic trap. While the approach presented in the previous section
does not apply here, because $P(\r,\theta)$ is not factorized, some
progress is still possible as we show below. The Stokes velocities of SPPs
are now position-dependent and read ${\bf v}_s=-\kappa \r/\zeta$,
where $\r$ is measured from the center of the trap. The physics is
very different from the sedimentation case since the particles are
effectively confined to $r<r_T\equiv v/\kappa$; At $r= r_T$ the trapping and
propulsive forces compensate exactly, hence preventing the particles
from moving further outside.

\begin{figure}
  \begin{center}
    \includegraphics{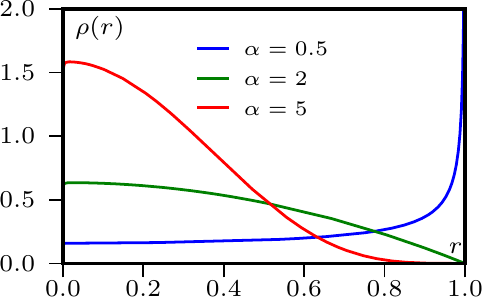}  \includegraphics{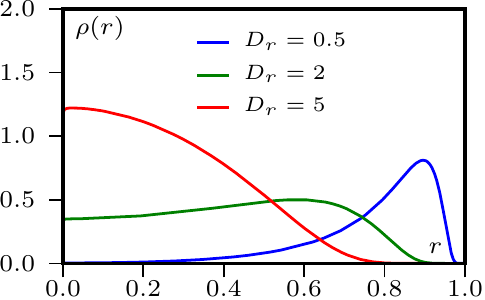}\\
    \includegraphics{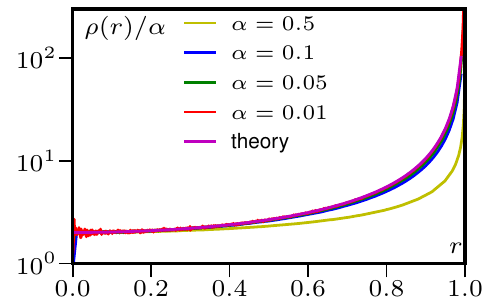}  \includegraphics{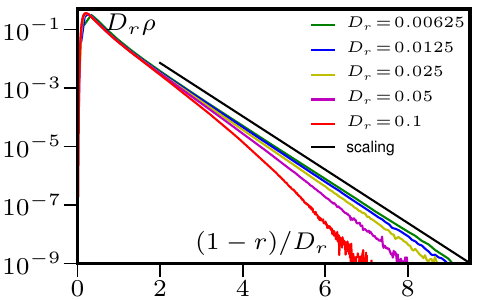}\\
  \end{center}
  \caption{SPPs in a harmonic trap ($v=\kappa=1$ in simulation
    units). As $\Phi=\kappa\tau$ increases, $\rho(r)$ goes from a
    Gaussian centered on $r\simeq 0$ to a distribution peaked at
    $r\simeq v/\kappa$; this holds for both RTPs (top left) and for
    ABPs (top right). As $\Phi$ diverges, the density of RTPs in the
    bulk scales as $1/\Phi$ (bottom left), while the density of ABPs
    in the bulk scales exponentially with $D_r$, and the profile
    $\rho(r)$ decreases exponentially with the distance to the trap
    boundary (bottom right). }
  \label{fig:traps}
\end{figure}

The effective equilibrium regimes correspond to $\tau\ll \lambda^{-1}$ and,
as expected, are the same for ABPs and RTPs, with
Boltzmann weights for the steady-state distribution $\rho(\r)$ (see
Fig.~\ref{fig:traps})
\begin{equation}
  \rho(\r)\propto  \exp\Big(-\frac{\kappa\, \r^2}{2kT_{\rm eff}}\Big);\qquad k T_{\rm eff}=\frac{v^2 \tau\zeta}{2}
\end{equation}
As the ratio $\Phi\equiv\kappa \tau$ between the persistence-length $v
\tau$ and the trap radius $r_T=v/\kappa$ increases, the time needed
for the particles to cross the trap becomes much smaller than the mean
reorientation time. Particles thus accumulate at the outskirts of the
trap where, on average, they point outward radially. How long they
persist in this state depends on the dynamical details; in
consequence, the density distributions of RTPs and ABPs differ in this
regime, particularly in the central region of the trap. Note that,
once again, this scenario differs from the fluctuating speed model
of~\cite{Szamel2014PRE} where the steady-state in a harmonic trap is
always a Gaussian, with fast enough particles reaching arbitrarily
large distances and slow enough particles remaining in the center of
the trap.

\paragraph{RTP in a harmonic trap.}

RTPs pointing outward at the trap boundary undergo instantaneous
tumble events at rate $\alpha$ whereafter they move across the trap
along a certain path. Despite the trap force, and somewhat
surprisingly, we show below this path to be a straight line segment,
which fully crosses the trap for large $\Phi$; In this limit, the
trajectory is thus a succession of chords of the circle in 2D. The
fraction of time spent by a particle in the bulk then corresponds to
the fraction of its run-time needed to reach the other side of the
trap; this time is proportional to $1/\Phi$. (This scaling is correct
although in practice the path is not followed at constant speed.)

Beyond the scaling $\rho(r)\propto \alpha/\kappa$ in the bulk, one can
furthermore compute a limiting shape of $\rho(r)$ as $\alpha/\kappa
\to 0$. Let us consider a particle at position $\r_0$ on the trap
boundary which tumbles to make an angle $\theta$ with the horizontal
axis. Its trajectory until the next tumble is simply given by
\begin{equation}
  \label{eq:trafABPtrap}
\r(t)=\r_0 e^{-\kappa t} +\frac{v}{\kappa} (1-e^{-\kappa t}) (\cos\theta,\,\sin\theta)
\end{equation}
which generalizes the intuitive result $r = (v/\kappa)(2 \exp[-\kappa
t]-1)$ for $\r_0=(v/\kappa,0)$ and $\cos \theta = -1$. (The latter
describes overdamped relaxation across a diameter of the circle to a
new equilibrium position opposite the previous one, and applies in the
case where the tumble exactly reverses the swim direction.)
Eq.~\eqref{eq:trafABPtrap} describes a linear trajectory joining
$\r_0$ to $\frac{v}{\kappa}u(\theta)$, where the trap force balances
again the propulsion force. The angular momentum
$L(t)=(\r(t)-\r_0)\wedge \dot \r(t)$ about the initial position $\r_0$
satisfies $L(0)=0$ and $\dot L=-\kappa L$ so that $L(t)=0$. The
velocity $\dot \r$ and $\r-\r_0$ are thus always parallel: the torques
about $\r_0$ exerted by the trap force and the propulsion force
balance to produce a straight trajectory at an angle intermediate
between $\u$ and $\r_0$; Only the speed $|\dot r(t)|\propto
\exp(-\kappa t)$ is thus varying.

\begin{figure}
  \begin{center}
    \includegraphics{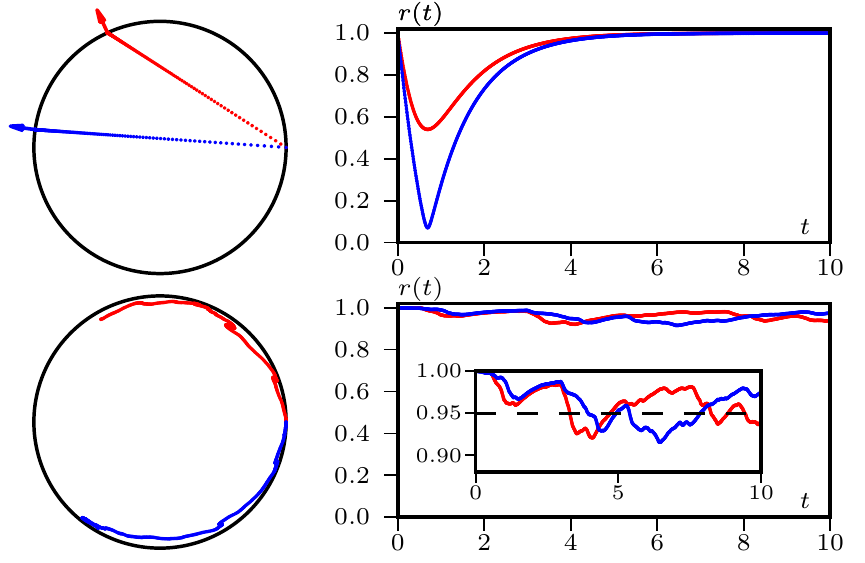}
  \end{center}
  \caption{SPPs in a harmonic trap ($\alpha=D_r=0.1$, $v=\kappa=1$ in
    simulation units). Left and right plots correspond to real-space
    trajectories and the evolution of $r(t)$, respectively. The arrows
    in the top left panel show the orientations of the particles. Top
    and bottom correspond to RTPs and ABPs, on similar time-scales. The dashed line in the bottom right panel correspond to $r= v [1-D_r/(2\kappa)]/\kappa$.}
  \label{fig:trajs}
\end{figure}

Let us now consider $\r_0=(v/\kappa,0)$. As shown in
Fig.~\ref{fig:trajs}, the distance to the center of the trap first
decreases from $r\sim v/\kappa$ to $r_{\rm
  min}(\theta)=v|\cos\frac{\theta}2|/\kappa$, which is reached after a
time $t=(\log 2)/\kappa$. The distance then increases again until
(almost) reaching $r\sim v/\kappa$ before the next tumble
happens. (Since the speed $|{\bf v+v}_s|$ vanishes as $r\to v/\kappa$,
the particle never reaches exactly the trap boundary.) The trajectory
thus crosses twice any annulus of radius $r>r_{\rm min}(\theta)$
within the trap, where it spends a fraction of its duration $\propto
\alpha (1/|\dot \r [t_+(\theta)]|+1/|\dot \r[t_-(\theta)]|)$. Here, we
have assumed that the total duration of the trajectory is $1/\alpha$
and $t_\pm(\theta)$ are the times at which the particle reaches
$|\r|=r$. These times satisfy
\begin{equation}
  (v^2-\kappa^2r^2)  \exp(\kappa t_\pm)=v^2 (1-\cos\theta) \pm \sqrt{2 \lambda^2 (1-\cos\theta)(r^2-r_{\rm min}^2)}
\end{equation}
so that the probability of finding a particle at position $r$ can be
computed as
\begin{equation}
  \label{eq:rhortptheory}
  \rho(r) = Z \frac{1}{2\pi} \int_{2\arccos(\kappa r/v)}^{2\pi-2\arccos(\kappa r/v)} \left(\frac{\alpha}{|\dot \r(t_1)|}+\frac{\alpha}{|\dot \r(t_2)|}\right) d\theta
\end{equation}
Note that $\rho(r)$ formally diverges as $r\to v/\kappa$; in practice
this is cut off because the finite duration ($\sim 1/\alpha$) of the
trajectory ensures that no particle exactly reaches the maximal radius
$r=v/\kappa$ at which its net speed vanishes. The smaller $\alpha$,
the closer to the trap boundary this cut-off takes place. We have
taken $\alpha$ out of the normalization $Z$ to make the scaling of
$\rho(r)$ apparent; $Z$ is then independent of $\alpha$ and equals
$Z\simeq \pi/2$. The bottom-left quadrant of Fig~\ref{fig:traps} shows
a very good agreement between Eq.~\eqref{eq:rhortptheory} and
simulations of RTPs for $\alpha/\kappa \leq 0.1$.

\paragraph{ABP in a harmonic trap.}

Contrary to RTPs, the orientation of ABPs diffuse slowly so that, as
$\theta$ varies, they simply slide along the boundary of the trap,
rarely visiting its inner region. This was used
recently~\cite{Yaouen,Yaouen2} to compute the distribution of ABPs
along the boundary of small traps of arbitrary shapes, assuming that
the particles effectively never leave the boundary of the trap. As we
show below, this asumption is a sound one since the density
of particles decreases exponentially as one moves away from the boundary
and is exponentially small in
$\Phi$ at the trap centre. Introducing the angle $\varphi$ between the direction of the
particle and the normal to the trap ${\bf e}_r$, the dynamics in the
$(r,\varphi)$ variables become
\begin{equation}
  \label{eqn:dynrphi}
  \dot r =-\kappa r + v \cos\varphi\qquad\dot \varphi=\sqrt{2D_r} \eta(t) -\frac{v}{r}\sin\varphi
\end{equation}
where $\eta(t)$ is a Gaussian unit white noise. Using
dimensionless variables $\tau=t D_r$, $\tilde r=r\kappa/v$, and
$\tilde \eta(\tau)=\eta(t)/\sqrt{D_r}$, Eqs.~\eqref{eqn:dynrphi} become
\begin{equation}
  \label{eqn:dynrphidl}
  \dot {\tilde r} =-\frac{\kappa}{D_r}(\tilde r - \cos \varphi)\qquad\dot \varphi=\sqrt{2} \tilde \eta -\frac{\kappa}{\tilde r D_r}\sin\varphi
\end{equation}
where $\tilde \eta(\tau)$ is also a Gaussian unit white noise. The
angle $\varphi$ thus undergoes rotational diffusion in an effective
potential $\frac{\kappa}{\tilde r D_r} (1-\cos\varphi)$ whose
amplitude diverges as $D_r/\kappa\to 0$. In this limit, $\varphi$
oscillates around $\varphi=0$ with Gaussian fluctuations $\langle
\varphi^2 \rangle=D_r/\kappa$.  Consequently, $\tilde r\simeq
1-D_r/(2\kappa)$: the particle is almost always at the border of the
trap, although the fluctuations of $\varphi$ prevent it from reaching
exactly $r=v/\kappa$ (see Fig.~\ref{fig:trajs}). The density of
particles in the bulk of the trap is thus much smaller than in the RTP
case since in a time $t\sim 1/D_r$, the particles are exponentially
unlikely to escape the trap boundary while RTPs would typically cross
the trap in times $t \sim 1/\alpha$. (Recall that these two times are
interchangeable in the effective equilibrium regime.)  The occurrence
of an effective potential, preventing particles from diffusing in
angle away from the outward normal direction, at first seems to
violate the fact that the dynamics of ABPs is torque-free. Note
however that there is no real potential for the angle $\theta$; the
effective potential for $\varphi$ arises not because there is a
torque, but because the translational motion of a particle along the
trap boundary naturally leads it towards a location where the outward
normal is parallel to the current orientation, hence leading to
$\varphi\simeq 0$. This effect is also present for RTPs but the discrete
angular dynamics effectively allows the RTP to immediately escape the
reorientation potential.

To reach the bulk part of the trap, Eq.~\eqref{eqn:dynrphidl} tells us
that $\varphi$ needs to climb the effective potential to turn and face
its destination. In principle, there are many stochastic paths leading
from the boundary to some inner target point at distance $r$ from the
center. As $\kappa/D_r\to 0$, however, the transition probability is
dominated by the most probable path leading to such a
point~\cite{Tailleur2008JPA} (an instanton). The probability of this path will be
given by $\exp \kappa \Delta E/D_r$, where $\Delta E(r/r_T)$ is a
geometric function depending on the exact location of the target. Since
reaching the center of the trap requires a fluctuation of $\varphi$
much larger than any required to reach another point close to the trap boundary, the
energy barrier increases as the distance of a target from the boundary
increases. In practice, a linear function $\Delta E(r/r_T)$ fits very
well the effective energy barrier:
\begin{equation}
  \rho({\bf r})\simeq \frac{C}{2\pi r_T^2} \Phi \exp\big[C \Phi (r/r_T-1)\big]
\end{equation}
where $C$ is a dimensionless constant. The density in the center of
the trap indeed decreases much faster than for RTPs, as
$\Phi\exp(-C\Phi)$. Somewhat surprisingly, this simple 
argument gives reasonably good agreement, with a geometric factor $C$
whose fit yields $C\simeq 2.15$, with the bulk density measured in
numerical simulations (See Fig.~\ref{fig:traps}).

  \subsection{Effective equilibrium for generic potentials}
  In this section we show that for generic potentials $V_{\rm ext}$,
  sufficient conditions to observe an effective equilibrium regime are given by
\begin{equation}
  \label{eqn:hypEE}
  |\xi^{-1} \grad V_{\rm ext}(\r)| \ll v;\qquad\text{and}\qquad |\xi^{-1} \Delta V_{\rm ext}(\r)|\ll \alpha,\,D_r(d-1)
\end{equation}
To show this, let us consider a system for which~\eqref{eqn:hypEE}
holds. The fluctuating hydrodynamics for ABPs and RTPs in the presence
of external drifts derived in~\cite{CT} reads:
\begin{eqnarray}
  \dot \rho &= -\frac \Omega d \grad (v {\bf p}) + \grad \cdot [\xi^{-1} \grad V_{\rm ext} \rho] + \grad (D_t \grad \rho)\label{eqn:EEGP1}\\
  A p_a &= -\grad_a (v \rho/\Omega) + \grad_b (\xi^{-1} \grad_b V_{\rm ext}p_a)+{\cal O}(\grad^2)\label{eqn:EEDP2}
\end{eqnarray}
where $\Omega$ is the surface of the unit sphere in $d$ dimensions and
$A$ equals $\alpha$ for RTPs and $D_r(d-1)$ for ABPs.  This
description is valid at times $t\gg A^{-1}$ and for fields $\rho,{\bf
  p}$ whose gradients are small on the scale of the persistence length
of SPPs, $|\grad \rho|/\rho,\,|\grad \p|/|\p|\ll A/v$; its derivation
in the isotropic case is the subject of section~\ref{mbp}, whose
notation differs slightly through introduction of a quantity $\varphi
= \rho/\Omega$.

Expanding the second term of the r.h.s of Eq.~\eqref{eqn:EEDP2} then yields
\begin{equation}
  (A -\xi^{-1} \Delta V_{\rm ext}) p_a = -\grad_a (v \rho/\Omega) + \xi^{-1} \grad_b V_{\rm ext}\cdot \grad_b p_a+{\cal O}(\grad^2)\label{eqn:EEDP3}
\end{equation}
Equation~\eqref{eqn:EEDP3} thus shows that $p_a\simeq {\cal
  O}(\grad)$. Using~\eqref{eqn:hypEE}, the Laplacian of $V_{\rm ext}$
can then be neglected and~\eqref{eqn:EEDP3} simplifies into
\begin{equation}
  p_a = -\frac{v}{A}\grad_a (\rho/\Omega) + {\cal O}(\grad^2)\label{eqn:EEDP4}
\end{equation}
The dynamics of $\rho$ is then given by
\begin{equation}
  \dot \rho = \grad \cdot \Big[\big(D_t+\frac{v^2}{A}\big) \grad \rho + \xi^{-1} \grad V_{\rm ext} \rho\Big ] \label{eqn:EEGPEE}
\end{equation}
which is nothing but the equilibrium Fokker-Planck equation of a
colloid of mobility $\xi$ subject to an external potential $V_{\rm
  ext}$, at a temperature
\begin{equation}
  kT_{\rm eff}=\xi \big[D_t+\frac{v^2}{A}\big]
\end{equation}
The steady-state solution of this equation is the Boltzmann weight
$\rho(x)\propto \exp[-\beta_{\rm eff} V_{\rm ext}]$, which satisfies
\begin{equation}
  \left|\frac{v}{A} \frac{\grad\rho}{\rho}\right|=\left|\frac{v}{A} \frac{\xi^{-1}\grad V_{\rm ext}}{D_t+\frac{v^2}{A}}\right|<\left|\frac{\xi^{-1}\grad V_{\rm ext}}{v}\right|
\end{equation}
which is indeed small, thanks to~\eqref{eqn:hypEE}. 

The two conditions~\eqref{eqn:hypEE} are enough to show the long
time dynamics of the SPP to amount to the equilibrium dynamics of a
colloid subject to an external potential $V_{\rm ext}$. Physically,
they require that the Stokes speed $v_s=-\grad V$ be much smaller that
the self-propulsion speed $v$ of the SPP, but also that its variations
over a run-length $ \grad v_s . v/A$ be much smaller than $v$.

\section{Fluctuating Hydrodynamics of ABPs and RTPs}
\label{mbp}
\label{sec:CG}
Following~\cite{CT}, we address a model using a general angular
relaxation dynamics, combining smooth angular diffusion with
diffusivity $D_r$ with pointwise random reorientations (tumbles) at
rate $\alpha$. The model has ABPs as one limit and RTPs as another. An
intrinsic translational diffusivity $D_t$ is also allowed for; the
swim speed $v$ is allowed to be position-dependent in the first part
of the derivation, $v(\r)$, and we will then turn to density-dependent
$v(\rho)$.  For a single particle, the probability density
$\psi(\r,\u,t)$ of finding the particle at position $\r$ moving in
direction $\u$ obeys exactly (in $d=2,3$)
\begin{equation}\label{EoM1}
  \begin{aligned}
    &\dot\psi(\r,\u) = - \n\cdot [v\u\psi(\r,\u)]+{\n_{\bf u} \cdot [D_r \n_{\bf u} \psi(\r,\u)]}\\
    &\quad+\nabla\cdot (D_t \nabla \psi(\r,\u))-\alpha \psi(\r,\u)
+\frac\alpha\Omega \int  \psi(\r,\u')\dO'
  \end{aligned}
\end{equation}
where $\n_{\bf u}$ is the {rotational gradient} acting on ${\bf u}$
and the integral is over the unit sphere $|{\bf u'}|=1$ of area
$\Omega$. The first term {on the right} is the divergence of the
advective current resulting from self propulsion and the last two
terms are loss and gain due to tumbling out of and into the
direction ${\bf u}$. The second and third terms account for rotational
and translational diffusion.

\subsection{Coarse-graining procedure}

The first part of the derivation below consists in using a moment
expansion to show that at large time and space scales, the
dynamics~\eqref{EoM1} amount to a Langevin equation, whose drift and
diffusion terms can be fully characterised. Using It\=o calculus, we
will then start from this single-particle Langevin dynamics to derive
a collective Langevin dynamics of the density field of $N$
non-interacting active particles. Since all the computations will be
done allowing for dependence of the microscopic parameters $v$,
$\alpha$, $D_r$ on the spatial position $\r$ of the active particle,
our computation applies to the case where these dependencies occur
through the density field $\rho(\r)$. This derivation will thus
provide the stochastic dynamics for the density field of $N$ interacting
active particles. Were we to consider purely ABPs, one could directly
use It\=o calculus to construct the dynamics of the density field, as
was done in~\cite{Farrell2012PRL,Romanczuk2012IF,Peruani2014JSM}. In the
general case,
it is the presence of the tumbles that makes it necessary to take a first diffusive
limit. Note that in lattice models, field theoretic methods can
be used directly that bypass this limitation and give the same final
result~\cite{Thompson2011JSM}.

\subsubsection{Moment expansion}
We first decompose $\psi$ as
\begin{equation}\label{decomp}
\psi(\r,\u,t) = \varphi + \p.\u +{\bf Q}:(\u\u-{\bf I}/d)+ \Theta[\psi]
\end{equation}
Here, $\varphi,\p,\Q$ are functions of $(\r,t)$, which parameterize
the zeroth, first and second angular ($d=2$) or spherical ($d=3$)
harmonic components of $\psi$, while $\Theta$ projects onto the higher
harmonics.  Note that in 2 dimensions, the angular harmonics are Fourier sine
and cosine series in $\theta$; clearly 
\begin{equation*}
  {\bf p . u} = p_1 \sin\theta+p_2\cos\theta\quad\text{and}\quad {\bf Q}\!:\!(u_iu_j-\frac{\delta_{ij}}2)=\frac{Q_{11}-Q_{22}}2 \cos2\theta+\frac{Q_{12}+Q_{21}}2\sin2\theta
\end{equation*}
span the subspaces generated by the first and second harmonics,
respectively. The same holds in $d=3$, where the components of any unit
vector $u_i$ are linear combinations of $Y_1^m, m = -1,0,1$ whereas
the components of the traceless tensor $u_iu_j-\delta_{ij}/3$ are
linear combinations of $Y_2^m, m = -2...+2$.

We now proceed order by order in the harmonics. Integrating Eq.~\eqref{EoM1} over $\u$ gives
\begin{equation}
\dot\varphi = -\frac 1 d\nabla(v {\bf p}) + \n (D_t \n \varphi) \label{EoM2}\\
\end{equation}
whereas multiplying eq.~\eqref{EoM1} by {\bf u}
and then integrating over $\u$ gives
\begin{equation}
  \begin{aligned}
\label {EoM3}
\dot p_a = -\n_a( v \varphi) - \big(D_r(d-1)+\alpha\big)
p_a + \n (D_t \n p_a)-\frac{2}{d+2} \nabla_b [v Q_{ab}]
  \end{aligned}
\end{equation}
Finally, multiplying eq.~\eqref{EoM1} by $\u \u -{\bf I}/d$ and
integrating over ${\bf u}$ yields
\begin{equation}
  \begin{aligned}
    \dot Q_{a b} =  - \frac{d+2}2 B_{a bcd} \n_c v p_d  -(2 d D_r+\alpha)  Q_{ab} + \n \cdot [D_t \n Q_{ab}] -\n_c v\chi_{abc} \label{EoM4}
  \end{aligned}
\end{equation}
where
$B_{abcd}=(\delta_{ac}\delta_{bd}+\delta_{ad}\delta_{bc}-
2 \delta_{ab} \delta_{cd}/d)/(d+2)$ and
 $\chi_{abc}$, which comes from higher order harmonics, will not play
any role in the following. The derivations of
Eqs.~(\ref{EoM2}-\ref{EoM4}) are detailed in Appendix~\ref{app:moments}.

\subsubsection{Diffusion-drift equations}
\label{sec:DDequations}
So far, beyond the assumed isotropy of $v(\r), D_{r,t}(\r)$ and
$\alpha(\r)$, no approximation has been made;
{Eqs}.~(\ref{EoM2}-\ref{EoM4}) are exact results for the time
evolution of the zeroth, first and second harmonics of
$\psi(\r,\u,t)$.  We then note that Eq.~\eqref{EoM2} is a mass
conservation equation:
\begin{equation}\label{DDdef}
\dot\varphi = -\n\cdot \J\quad\text{with}\quad \J = \frac{1}{d} v \p - D_t\nabla\varphi
\end{equation}
with a current $\J$ which involves the first harmonic $\p$. We will now
use a large time and space scale limit to obtain $\J$ as a function of
$\varphi$. 

We first note that $\varphi$ is locally conserved and is thus a slow
mode: the relaxation of a density perturbation on a scale $\ell$
occurs in a time that diverges as $ \ell\to\infty$. On the contrary,
$\p$ and $\Q$ are fast modes, whose relaxation rates are given by
$t_\p=\alpha+D_r(d-1)$ and $t_\Q=\alpha+2 d D_r$. For times greater
than $t_{\p,\Q}$, one thus assumes
$\dot\p=\dot\Q=\Theta[\dot\psi]=0$. Eqs.~\eqref{EoM3} and \eqref{EoM4}
then give the ordinary differential equations that $\p$ and $\Q$
satisfy quasi-statically as $\varphi$ evolves on time-scales much
larger than $t_{\p,\Q}$. By itself, this creates a non-local
constitutive relation between $\J$ and $\psi$ which still involves all
harmonics.

Next we explicitly carry out a gradient expansion, yielding for $\Q$:
\begin{equation}
Q_{ab} = -\frac{1}{2 d D_r +\alpha
}\left(\frac{d+2}{2}B_{abcd}\nabla_c(vp_d)+\nabla_c( \chi_{abc})\right)+{\cal O}(\n^2)
\end{equation}
from which the quasi-stationary $\p$ then follows via~\eqref{EoM3} as
\begin{equation} \label{pgrad}
\p = -\frac{1}{(d-1)D_r+\alpha} \n(v\varphi) + O(\n^2)
\end{equation}
At this order in the gradient expansion, closure is achieved without
needing further information on harmonics beyond the first; the
current $\J$ is given by
\begin{equation}\label{DD}
 \J = -\frac{v}{d(d-1)D_r+d\alpha}  \n(v\varphi)-D_t \n\varphi
\end{equation}
At this order, Eq.\eqref{DDdef} corresponds
to a diffusion-drift approximation of the microscopic master
equation~\eqref{EoM1}, given by:
\begin{equation}\label{VD}
  \dot\varphi=-\n\cdot [-D \nabla \varphi + \V \varphi]
\end{equation}
where the diffusivity and drift velocity obey
\begin{equation}\label{eqn:DVdef}
D = \frac{v^2}{d(d-1)D_r+d\alpha}+D_t \quad\text{and}\quad \V = \frac{-v\n v}{d(d-1)D_r+d\alpha} 
\end{equation}
At this level of description, though not for the exact results that
preceded it, RTPs and ABPs are seen to be equivalent. That is,
in~\eqref{VD}, the tumble rate $\alpha$ and rotational diffusivity
$D_t$ enter only through the combination $(d-1)D_r+\alpha$, so that
the two types of angular relaxation are fully
interchangeable. Conversely, given a sequence of snapshots showing the
large-scale evolution of the local densities of SPPs, one cannot
determine whether the system is composed of ABPs or RTPs. Note that
``large scale" here excludes some of the trap problems considered in
Section \ref{extpot} where the confinement length is smaller than the
persistence length of the self-propelled motion.

\subsubsection{Many-body physics}
Eqs.~\eqref{DDdef} and \eqref{VD} give the evolution for the
probability density of one particle at diffusion-drift level.
Following \cite{TC} we can now consider an assembly of particles whose
motility parameters $v,\alpha,D_r$ and $D_t$ depend on position both
directly and indirectly, through a functional dependence on the
microscopic density $\rho(\r,t) \simeq \sum_\mu\delta(\r
-\r_\mu(t))$. Since we are considering a multiplicative noise whose
variance is a functional of the density field $\rho$, the construction
of the Langevin equation associated to $\rho$ is technically more
involved than in the additive case~\cite{Dean}. First, we go
from the Fokker-Planck equation~\eqref{VD} to the equivalent
It\=o-Langevin equation for an individual particle position
$\r_\mu(t)$:
\begin{equation}\label{eqn:Langmicro}
  \dot \r_\mu(t)= \V + \n_{1} D(\r_\mu,[\rho])+\big(\n_{\r_\mu} \frac{\delta}{\delta \rho(\r_\mu)}\big) D(\r_\mu,[\rho]) +\sqrt{2D}{\boldsymbol \eta}
\end{equation}
where $\n_{1}D(\r_\mu,[\rho])$ represents a gradient with
  respect to the \textit{first} argument of $D$ and not with respect
  to its implicit dependence on $\r_\mu$ through the field
  $\rho(\r')=\sum_\alpha \delta(\r'-\r_\alpha)$. On the contrary, the operator $\n_{\r_\mu}
  \frac{\delta}{\delta \rho(\r_\mu)}$ applies to the functional
  dependence of $D$ on the field $\rho(\r')$ and the subscript
  $\r_\mu$ shows $\grad_{\r_\mu}$ to apply to dependencies on $\r_\mu$ and
  not on $\r'$ or $\r_{\alpha\neq \mu}$ that may appear in $D$. (For
  more technical details, see Appendix~\ref{app:Itodrift}.)  The two
  terms $\n_{1} D(\r_\mu,[\rho])+\n_{\r_\mu}
\frac{\delta}{\delta \rho(\r_\mu)} D(\r_\mu,[\rho])$ form the
so-called ``spurious drift'' term due to the It\=o time
discretisation. (This term is, of course, not spurious, but necessary
once one uses the It\=o convention to define that
discretization~\cite{Risken,oksendal}.)  The atypical form of this
spurious drift is due to $D$ depending on $\r_\mu$ both explicitly,
which explains the first term, and implicitly---through $\rho(\r')$---
which explains the second one. In Appendix~\ref{app:Itodrift}, we
show that the density $\rho(\r,t)$ then obeys the many-body Langevin
equation \cite{TC}
\begin{equation} \label{coll}{
    \dot\rho = -\n.\left(\V[\rho]\rho - D[\rho]\nabla \rho+\big(\n_\r \frac{\delta}{\delta \rho(\r)}\big) D(\r,[\rho]) + (2D\rho)^{1/2}{\bf \Lambda}\right)
}
\end{equation}
with white noise $\langle \Lambda_i(\r,t)\Lambda_j(\r',t')\rangle =
\delta_{ij}\delta(\r-\r')\delta(t-t')$. The gradients
  in~\eqref{coll} have no subscript `$1$' since there are no more
  ambiguities at this stage: the derivatives are taken with respect to
  the spatial coordinate $\r$, which does not enter $\rho(\r')$, and not
  with respect to the position of one of the $N$ particles as in~\eqref{eqn:Langmicro}. Note that
when the diffusivity of particle $\mu$ does not include
`self-interaction', i.e. $D=D(\r_\mu,[\rho-\delta(\r-\r_\mu)]$ or is
the result of a convolution between $\rho(\r')$ and a symmetric kernel
$K(\r)$, i.e. $D=\int \d\r' K(\r_\mu-\r')\rho(\r')$, then $\n_{\r_\mu}
\frac{\delta}{\delta \rho(\r_\mu)} D(\r_\mu,[\rho])=-\n K(0)$ vanishes
(see Appendix~\ref{app:Itodrift} for details) and one recovers the
more standard equation
\begin{equation} \label{collsimp}{ \dot\rho = -\n.\left(\V[\rho]\rho -
      D[\rho]\nabla \rho+ (2D\rho)^{1/2}{\bf
        \Lambda}\right) }
\end{equation}
The functionals $v[\rho]$, $\alpha[\rho]$ and $D_{t,r}[\rho]$
in~\eqref{eqn:DVdef} then define for the interacting particle system the
many-body drift velocity and diffusivity $\V[\rho]$ and $D[\rho]$ for
use in Eq.~\eqref{coll}.

Starting from the many-body It\=o-Langevin equation~\eqref{coll}, one can then
derive a functional Fokker-Planck equation for the evolution of the probability
density $P[\rho(\r),t]$ of finding the system with a density field
$\rho(\r)$ at time $t$:
\begin{equation}\label{eq:FPErho}
  \dot P[\rho]=\int \d \r \frac{\delta}{\delta \rho(\r)} \n\cdot \left[\rho \V -D \n \rho -D\rho \left(\n_{\r}\frac{\delta}{\delta\rho(\r)}\right)\right]P[\rho]
\end{equation}
The technical details, which show the importance of the atypical
spurious drift, are detailed in appendix~\ref{app:FP}.

\subsubsection{Connection to large-deviation functionals; role of
    noise} 
\label{sec:LDF}
A crucial observation, first made for RTPs in \cite{TC},
is that~\eqref{coll} reduces, under specific conditions, to a
description of passive Brownian particles (PBPs) with a specified free
energy functional $\beta\calf[\rho] = \beta\calf_{\rm ex}[\rho] +
\int\rho(\ln\rho-1)dx$.  (In what follows we use thermal units in
which $\beta = 1$.) This is best seen by noting that the Fokker-Planck
equation~\eqref{eq:FPErho} admits flux-free solution
\begin{equation}\label{eqn:ffsolution}
  \left[\rho \V -D \n \rho -D\rho \left(\n_\r\frac{\delta}{\delta\rho(\r)}\right)\right]P[\rho]=0
\end{equation}
whenever there exists a functional ${\cal F}_{\rm ex}[\rho]$ which
satisfies the condition
\begin{equation}
\V([\rho],\r)/D([\rho],\r) = -\nabla_\r(\delta \calf_{\rm ex}[\rho]/\delta \rho(\r)) \label{integrable}
\end{equation} 
Indeed, in such a case $P[\rho]=\exp[-\beta {\cal F}[\rho]]$
satisfies~\eqref{eqn:ffsolution}. As was noted
in~\cite{Peruani2014JSM,Thompson2011JSM}, $s[\rho]=\beta {\cal F}/V$,
with $V$ the volume of the system, is the large deviation function of
the density profile $\rho$
\begin{equation}
  s[\rho]=-\lim_{V\to\infty} \frac 1 V \log P[\rho] 
\end{equation}
It is at first sight very surprising that one can compute this object
for any kind of interacting out-of-equilibrium system. However its
computability depends on \eqref{integrable} holding true, and indeed
this is not true in any general manner. It is however satisfied to
leading (zeroth) order in a gradient expansion of ${\cal F}_{ex}$.
If \eqref{integrable} holds,
the system is completely equivalent to an equilibrium system of
passive particles with excess chemical potential gradient $\delta
{\cal F}_{\rm ex}/\delta \rho(\r)$. This equivalence holds not only at
the level of the steady-state but also at the level of the
dynamics~\eqref{eq:FPErho}. This means that, at this macroscopic
level, the dynamics is effectively an equilibrium one and, for
instance, satisfy the Onsager-Machlup symmetry between excursion and
relaxation~\cite{Tailleur2008JPA}. However this does not mean that the
phenomenology of the system reduces to an equilibrium one. For
instance, the slowing down of active particles at high density causes
an effective many-body attraction between the equivalent passive
particles, characterized by a $\calf_{\rm ex}$ whose local part has
negative curvature in $\rho$. This kinetic slowdown can trigger a
phase separation mechanism, which we detail in the following section,
which would be impossible at equilibrium. (Making the viscous drag
on each particle an increasing function of local density would generate a similar kinetic slowdown
in an equilibrium system, but
the Boltzmann distribution would be insensitive to this effect.)

If the condition~\eqref{integrable} is not met, however, the dynamics
is not equivalent to PBPs with conservative interactions and is
therefore ``irreducibly'' active. As we will show later, the
equivalence established above will only hold for homogeneous systems
and, when the kinetic slow-down triggers a phase-separation,
higher-order gradient terms comes into play at the large-deviation
level, and break the mapping to equilibrium.

\subsubsection{Mapping to equilibrium}
\label{sec:MIPS-eq}
The simplest case is where $D_t=0$. Here the left hand side of~\eqref{integrable} is
$-\nabla \ln v[\rho]$ and we then require $\delta
\calf_{ex}[\rho]/\delta\rho(\r) = \ln(v([\rho];\r)$. 
The simplest first approach is to assume that the functional dependence of swim speed on density is strictly local, so that $v([\rho];\r) =
v(\rho(\r))$.  We then have $\calf_{ex} = \int f_{ex}(\rho(\r)) d\r$ where
$f_{ex} = \int_0^\rho\ln v(\lambda)d\lambda$. This structure in the free energy is equivalent to having a passive system whose chemical potential obeys $\mu = \ln(\rho v)$; the mean particle current then obeys 
\begin{equation}
{\bf J} = - \rho D \nabla \ln(\rho v) \label{chempotform}
\end{equation}
Noting that in our chosen units ($\beta =1$) the mobility coincides
with the diffusivity $D$, this is the expected form for the stated
chemical potential.  When ${D}_t\alpha $ is a nonzero constant, the result
for $f_{ex}$ generalizes to
\begin{equation}
f_{ex}
  = \half\int_0^\rho\ln[ v(\lambda)^2+ d {D}_t\alpha] d\lambda
\end{equation}
This case is explored in \cite{MIPS}.

To approximate the excess free energy by a local function is a widely
used approximation in equilibrium systems which makes sense at
(Landau) mean-field level where fluctuations are neglected. (If the
excess free energy were genuinely local, short-length scale
fluctuations would be out of control, so to go beyond mean field
theory requires treatment of nonlocality even if that is weak).  The
chosen form (with $D_t = 0$) leads to a spinodal instability whenever
{$dv/d\rho < -v/\rho$} \cite{TC,MIPS}. The system is then equivalent
to an equilibrium system undergoing liquid-gas phase separation due to
attractive forces and the spinodal instability is complemented by
binodals, which can be characterised analytically in some cases
(See~\cite{MIPS} for a recent review).

For ABPs with collisions, one finds empirically that $v(\rho)\simeq
v_0(1-\rho/\rho^*)$ with $\rho^*$ a near-close-packed density beyond
which self-propulsion is effectively arrested. The system is then
spinodally unstable for $\rho \ge \rho^*/2$
\cite{fily,sten1,sten2}. Stability is restored by packing constraints
at high enough density, but in the present approach (collisions
replaced by an effective density-dependent propulsion speed),
empirical corrections to $\calf_{ex}$ are needed to account for this
\cite{sten1}. An alternative is to impose a sharp cutoff so that $f =
+\infty$ for $\rho >\rho^*$; this is used in \cite{Solon2}.

The presence of a finite $D_t$, either due to Brownian motion or
resulting from the random collision of particles (in which case $D_t$
will depend on $\rho$), alters the shape of the spinodal curves which
instead obey $1+\rho v'/v<-d D_t/(v^2\tau)$. This causes the spinodals
to meet at a critical point; while in qualitative agreement with the
shape of the phase diagram reported for ABPs with repulsions (see,
e.g.~\cite{fily,sten1,redner}), this is somewhat accidental since the
actual role of $D_t$ in the simulated dynamics is negligible.  The
observed critical point is better viewed in terms of the competition
between slowing down at high density (promoting motility-induced phase
separation) and the buildup of particle mechanical pressure~\cite{Solon2}. The
latter stems primarily from the pair interaction, rather than the
small ideal gas part which is proportional to $D_t$
\cite{Brady,Solon2}.  Numerically, the spinodals are hard to locate
precisely, whereas binodals can be located by looking directly at the
coexisting densities in phase-separated states.  The point at which
phase separate is seen kinetically in ABP simulations often lies in
between the spinodal and binodal; it depends in general on nucleation
rates and is subject to large finite-size corrections.  Very large
scale simulations are thus required for accurate phase diagram
determination~\cite{sten1,sten2}.

\subsubsection{Beyond the local approximation}

When a system phase separates, large gradients develop and the
gradient expansion cannot be truncated at lowest order: the local
approximation to ${\cal F}$ no longer yields a good approximation to the
large deviation functional. In such a case, one
has to look for higher order gradient terms, which stem from two
different mechanisms. First, for real particles, the interactions
between particles are never perfectly local: even for hard-core repulsions,
the particle size defines a finite interaction range.
Second, we have dropped higher angular harmonic
contributions by appeal to a gradient approximation. Retaining higher
order gradients thus requires two different type of terms whose impact
on motility-induced phase separation we now discuss.

\paragraph{{Non-local $v(\rho)$.}}

The local form of the free energy can be used to predict the binodal
densities for phase coexistence via the the common tangent
construction on $f = f_{ex} + \rho(\ln\rho-1)$ \cite{TC}. In this
construction one equates the chemical potential $df/d\rho$ and the
`thermodynamic' pressure $\rho df/d\rho - f$ in the two phases; this
means that a single line can be drawn on a plot of $f(\rho)$ that is
tangent at the binodal densities and lies below $f$ everywhere
else. However, there is a hidden pitfall here: this construction
implicitly assumes that, whatever the nonlocal terms are, these
continue to obey~\eqref{integrable}. Whenever they do, a free energy
still exists and, so long as it does, the binodal densities do not
depend on the precise form of the nonlocal terms. Conversely, if a
free energy does not exist, then the binodal conditions can depend
explicitly on the nonlocal terms. This is explored in some detail,
within the simplification of a $\phi^4$ effective field theory, in
\cite{ModelB}. For ABPs, numerical evidence of the breakdown of the
common tangent construction was reported in \cite{sten1,sten2}.

The lowest order nonlocal theory can be constructed by assuming that the functional dependence of swim speed on density takes the form~\cite{TC}
\begin{equation}
v[\rho] = v(\hat\rho) \simeq v(\rho + \gamma^2\nabla^2\rho) \simeq v(\rho) + v'(\rho)\gamma^2\nabla^2\rho \label{taylor}
\end{equation}
This represents a quasi-local dependence on a quantity $\hat\rho$ that samples the local density isotropically in a region of size $\gamma$. For programmed slowing-down (e.g., quorum-sensing)
the simplest assumption is that $\gamma$ is a density-independent constant. However for collisional slowing down in ABPs, a better estimate is $\gamma\simeq \gamma_0v(\rho)/[(d-1)D_r+\alpha]$ which, with $\gamma_0$ of order unity, is the distance travelled during one angular relaxation time. Substituting the nonlocal form for $v$ in \eqref{chempotform} gives to leading order
\begin{eqnarray}
{\bf J} &=& - \rho D \nabla \mu \\
\mu & =&  \ln(\rho v) - \frac{\gamma^2}{v}v'\nabla^2\rho \label{chempotform2}
\end{eqnarray}
Only if $\gamma^2v'/v$ is a constant, independent of density $\rho$,
does this form of chemical potential support the existence of a free
energy. If this combination is constant -- such as for the case where
$\gamma$ is constant and $v(\rho) \sim \exp[-a\rho]$~\cite{TC} -- then
the common tangent can legitimately be used to predict the binodal
densities. In all other cases it cannot be relied upon
\cite{sten1,ModelB}. Note that in practice, only quantitative
differences with the local theory have been noticed when simulating
models with such non-local $v([\rho])$~\cite{sten1,ModelB}: the phase
separation still occurs, the coarsening law is not much affected, and
one only observes quantitative shifts of the binodals. Importantly,
the \textit{concept} of a binodal is maintained: the densities of
coexisting phases do not change as the global average density in the
system is varied between the two coexistence values.

\paragraph{{Other gradient terms.}}

The second source of higher order gradient terms is the gradient
expansion used to close the spherical harmonics expansion in
Section~\ref{sec:DDequations}. Were we to pursue this gradient
expansion to higher orders, we would obtain a set of ordinary
differential equations for $\p$, $Q$, etc. rather than the simple
algebraic relation~\eqref{pgrad} between $\p$, $\varphi$ and
$\n\varphi$. Solving these equations in terms of $\varphi$ and
reinjecting into $\J$ would then not lead to a simple Fokker-Planck
equation, from which we would not be able to derive a microscopic
Langevin equation like~\eqref{eqn:Langmicro}, which is the starting
point of our approach. The equivalence between ABPs and RTPs at this
higher order in gradients is thus questionable. This echoes the
fact that the difference between the two models becomes
more important at short length scales.
As we will show in Section~\ref{sec:simumicro},
however, these differences hardly impact the phase diagram.

For ABPs, since one directly starts at the microscopic level with
coupled Langevin equations, one can bypass the construction of the
microscopic Langevin equation~\eqref{eqn:Langmicro} and directly use
It\=o calculus to obtain a stochastic equation for the probability
density $\rho(\r,\u)$ of finding particles at $\r$ going in the
direction
$\u$~\cite{Farrell2012PRL,Romanczuk2012IF,Peruani2014JSM}. One then
has to project this equation onto successive harmonics and use an
appropriate truncation. Again, the equation~\eqref{coll} amounts to
second order gradient expansion. It is then easier to pursue this
development to higher orders, which would yield coupled stochastic
equations for $\rho$, $\p$, $Q$, etc.. As far as we are aware, this has
not be carried out in the literature for ABPs or RTPs, but progress
along this path can be found for aligning nematic
particles~\cite{Bertin2013NJP}.

For ABPs, a similar path has been followed at mean-field level, albeit
without a proper derivation of the noise terms, to arrive
at~\cite{fily,Henkes,Loewen2a}
\begin{equation}
\begin{aligned}
  \label{eqn:MF}
  \dot\rho&=D\Delta \rho - \nabla \cdot [v {\bf p}]\\
  \dot p&=-D_r {\bf p} -\frac 1 2 \nabla [v\rho] + D \Delta {\bf p} \\
\end{aligned}
\end{equation}
One can then always complete these equation by adding ad-hoc Gaussian
noise~\cite{fily} or by assuming that, were the coupled
equations~\eqref{eqn:MF} to be written in a gradient form (implying the existence of a free energy
functional) the noise
terms would be whatever is required to give an equilibrium like-dynamics. As we
have shown in Section~\ref{sec:LDF}, however, the derivation of the
noise terms is a crucial step to determine whether or not the system
really maps onto an equilibrium one. Indeed, the integrability
criterion~\eqref{integrable} is derived by requiring a flux-free
steady-state in the Fokker-Planck equation. This condition crucially
relies on the fact that the variance of the noise obeys
a fluctuation-dissipation relation with the mobility.

Nevertheless, since the derivations of these noise terms are particularly
difficult, a first strategy can be to look at whether the
equations~\eqref{eqn:MF} have the gradient structure of an equilibrium model. While analyzing the linear
stability of~\eqref{eqn:MF} for a linearly decreasing $v(\rho)$ reveals
a spinodal decomposition scenario very similar to the one described in
section~\eqref{sec:MIPS-eq}, it was explicitly shown in~\cite{Loewen2a}
that these equations are not of gradient form. However, expanding close to the
linear instability, the authors of~\cite{Loewen2a} were able to map
these equations onto a Cahn-Hilliard model, but with an effective free energy
which depends on where in the phase diagram the expansion is carried
out. This revealed a rather surprising feature: the addition of the
$\Delta{\bf p}$ terms breaks the global mapping to equilibrium but seems to
preserve a local one. This however fundamentally changes the structure of the
phase diagram: there is nothing to guarantee that the binodals stay in fixed
positions as one moves along the `tie line' between them (and without this property
the concept of a binodal is inapplicable). Moreover, according to this
calculation, a large part of the transition line is changed from a first order
to a second order transition, in contrast to the isolated critical point usually seen
in liquid-gas type phase-separation.

\begin{figure}
  \begin{center}
    \includegraphics{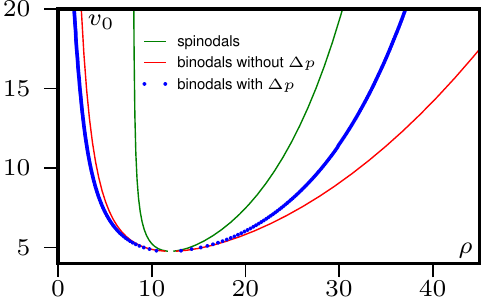}
    \includegraphics{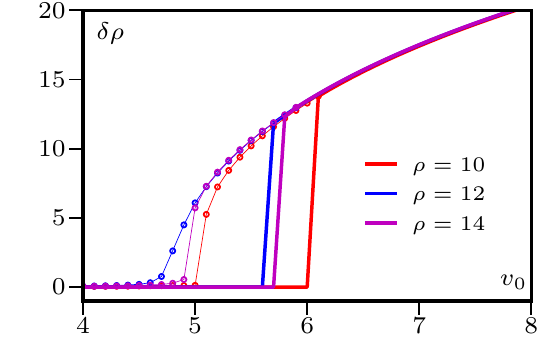}
  \end{center}
  \caption{The phase diagram ({\bf left}) predicted by the theory
    without the $\Delta{\bf p}$ term (red and green lines) are only
    qualitatively affected by the $\Delta {\bf p}$ term
    (blue). Hysteresis loops ({\bf right}), showing the first order
    nature of the transition, can be seen on approach to the critical
    point by ramping up (solid lines) or down (connected
    symbols). $D=0.25,\, v_1=0.25,\, \phi=4,\, \tau=1/D_r=1$.}
  \label{fig:Deltapterm}
\end{figure}

To test the predictions of this approach, we simulated directly the equations~\eqref{eqn:MF} using
spectral-methods and semi-implicit time-stepping. We chose a form of
$v(\rho)$ which does not lead to $v(\rho)<0$ at any density and thus
does not require an ad-hoc cut-off at the level of the effective free
energies:
\begin{equation}
\label{eqn:bestvofrho}
  v(\rho)^2=v_0^2+(v_1^2-v_0^2)(1-e^{-\rho/\phi})
\end{equation}
For this choice, we can compute analytically spinodals and
binodals predicted by the local theory~\cite{MIPS} and hence evaluate
precisely the impact of the $\Delta{\bf p}$ term. The scenario revealed by
our simulations does not show any of the surprising phenomenology
predicted by the methods of~\cite{Loewen2a}: the binodals are quantitatively shifted
by this new gradient term, in qualitative agreement with the effect of
gradient terms stemming from a non-local $v(\rho)$~\cite{ModelB}, but
the binodals
do not vary along the tie-line in the coexistence region (where the lever rule still applies).  
Also, the
transition line seems to remain first-order on approach to the
critical point, exhibiting the familiar hysteresis curves (see
fig.~\ref{fig:Deltapterm}). 

In principle, this outcome could depend on our choice of
$v(\rho)$; note that direct comparison with~\cite{Loewen2a} is not possible
since the theory developed there relies on the addition of confining
terms at the level of the free energy that do not exist at the level
of the PDE, Eq.~\eqref{eqn:MF}. However this interpretation seems unlikely given
the growing breadth of literature supporting the existence of well-defined
binodals in MIPS. An alternative possibility is that the new and unusual phenomenology
predicted from the approach of~\cite{Loewen2a} is an artefact of a quasi-linear, noiseless
treatement of what is in fact a noisy nonlinear transition. Further study is needed
to resolve this issue, for instance by carrying out the
analytical procedure in~\cite{Loewen2a} for the particular 
$v(\rho)$ chosen in~\eqref{eqn:bestvofrho}.

\begin{figure}
  \begin{center}
    \includegraphics{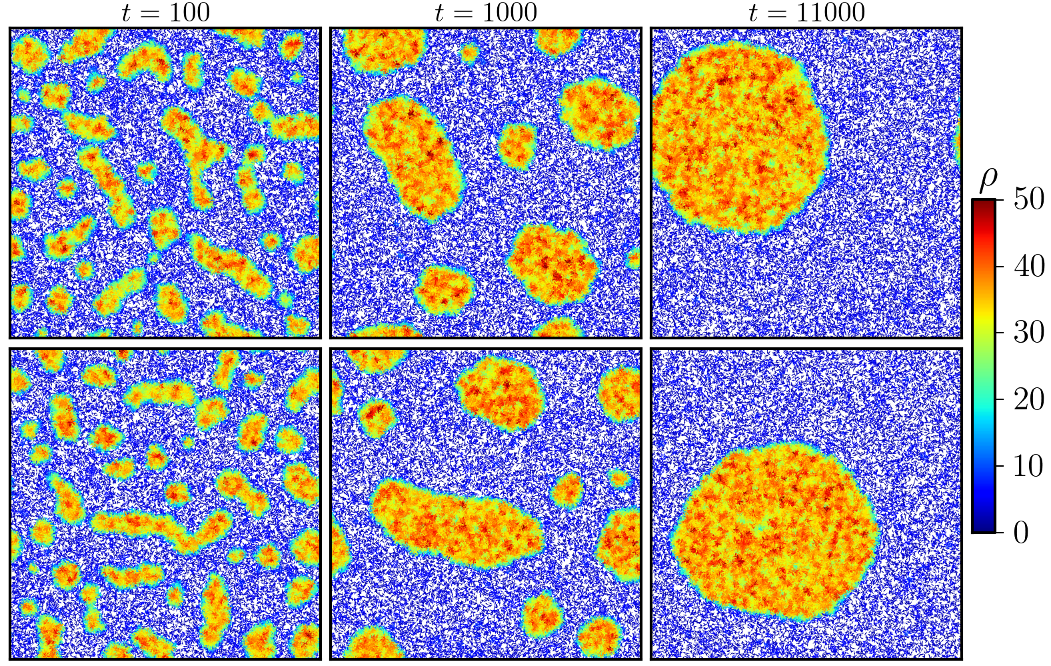}
  \end{center}
  \caption{Coarsening dynamics of RTPs (top) and ABPs (bottom)
    interacting via the density-dependent
    swim-speed~\eqref{eqn:bestvofrho} with $v_0=15$, $v_1=0.25$,
    $\varphi=4$, $\tau=1$, $D_t=0.25$. Simulated for $N = 120\,000$ particles
in a box of side $L = 100$.}
  \label{fig:ABPRTPvofrho}
\end{figure}

\begin{figure}[ht]
  \begin{center}
    \includegraphics{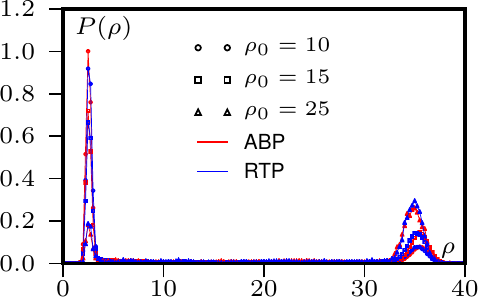}    \includegraphics{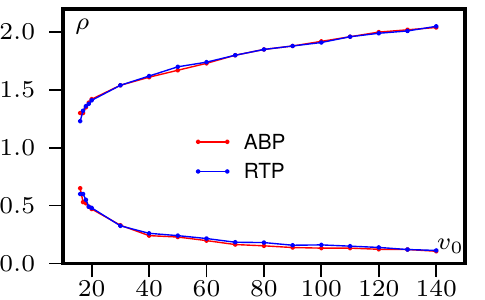}
  \end{center}
  \caption{{\bf Left}: The local density distributions $P(\rho)$ of
    ABPs and RTPs interacting via the density-dependent swim speed
    Eq.~\eqref{eqn:bestvofrho} with $v_0=15$, $v_1=0.25$, $\varphi=4$,
    $\tau=1$, $D_t=0.25$. The two systems overlap very well, showing
    that not only the mean coexistence densities in the two models but
    also their fluctuations are very similar. As expected, changing
    the mean density $\rho_0$ does not change the coexisting
    densities.  {\bf Right}: Phase diagrams of RTPs and ABPs
    interacting via WCA potential. The boundary lines are obtained by
    computing the binodals as a function of $v$ in the phase
    coexistence region for various global densities $\rho_0 = N/L^2$,
    to check that the binodals are indeed independent of
    $\rho_0$. Again, a surprisingly good overlap between the two
    systems is observed.}
  \label{fig:pdwca}
\end{figure}

\subsubsection{Comparison between microscopic simulations of RTPs and ABPs}
\label{sec:simumicro}
Since gradient terms can alter the equivalence between ABPs and RTPs,
we carried out extensive microscopic simulations of MIPS for both
models in two dimensions for $N$ particles in an $L\times L$ square
domain.  We first compared the two models in the case of a
density-dependent swim speed $v(\rho)$ given
by~\eqref{eqn:bestvofrho}. The coarsening of both phase separating
systems is shown in figure~\ref{fig:ABPRTPvofrho} and shows a
strikingly similar dynamics. Beyond the apparent similarity of the two
dynamics, one can compute the steady-state density distributions
$P(\rho)$ of the two systems in the phase separated region, which
shows that both the mean densities and their fluctuations in each
phase are also very similar (see Fig.~\ref{fig:pdwca}). Note that as
expected, the coexisting densities along the tie-line in the
coexistence region do not depend on the mean density $\rho_0=N/L^2$.

We then simulated both ABPs and RTPs in the case where slowing down is
caused by repulsive interactions (represented as a WCA potential as
in~\cite{fily,baskaran,sten1}). We know that the $v(\rho)$ theory
correctly captures the slowdown in these models but that it lacks the
direct interparticle forces which are responsible for saturating
density in the high-density phase~\cite{Solon2}. Therefore, the
dynamical equivalence of ABPs and RTPs at large scales might break
down for the repulsive case where much of the physics depends on
short-distance collisional events taking place below the
coarse-graining length scale of the fluctuating hydrodynamics theory.
Suprisingly however, the phase diagrams of repulsive ABPs and RTPs
collapse onto each other upon the usual rescaling $\alpha=(d-1)D_r$
(see Fig.~\ref{fig:pdwca}).  Thus, although we have shown that the
details of the angular dynamics are important for SPPs confined to
traps of size comparable to the run-length or smaller, the presence of
short-length scale physics in the mechanism for MIPS does not create a
similar dependence on details, as judged either by the qualitative
kinetic observations or a quantitative study of phase diagrams.

\section{Conclusions}
Motivated by a wish to understand for active systems the relation
between microscopic dynamics and macroscopic behaviour, we have in
this paper compared in detail two distinct but related classes of
self-propelled particles: Active Brownian particles and run-and-tumble
particles. These differ only in the details of their angular
relaxation (continuous versus discrete). We considered first
one-body-problems involving non-interacting particles in either
uniform fields (e.g., gravity) or isotropic harmonic traps. In both
cases, for weak enough forces there is an ``effective equilibrium"
regime in which the characteristic length scale set by the balance of
external and propulsive forces is large compared to the run length,
defined as the distance a free particle can travel during its
orientational relaxation time. In this regime, the two types of
dynamics (ABP and RTP) are equivalent modulo a simple mapping between
parameters, which corresponds to equating the angular relaxation time
in the two cases.  However, for stronger confinement the corrections
to this picture are specific to the chosen dynamics. In the case of
harmonic traps, this difference is particularly striking: in the
regime where the confinement length is large compared to the run
length, there is an exponentially strong suppression of the density of
ABPs at the trap centre when compared to RTPs of the same rotational
relaxation time. This can be attributed to a subtle but physically
simple mechanism whereby, in the strong trapping regime, particles
tend to remain a long time at the outer edge of the trap, where
external and propulsive forces balance. For RTPs this balance is
suddenly destroyed with each tumble event, whereas for ABPs, which
update their orientations gradually, there is a tendency to drift
around the perimeter so that the balance is continually maintained,
causing only an exponentially small escape rate from the surface
region. We also considered the effect of the microscopic rotational
dynamics on MIPs, or motility-induced phase separation. For this
process we gave a fuller presentation of results summarized in our
earlier papers concerning the derivation of collective equations for
the particle density field, from which criteria for MIPS are easily
found within a limiting approximation corresponding to neglect of
gradient terms in the large deviation functional. This criterion is
again insensitive to the choice of rotational dynamics. It is
explicitly derived from a model where particles interact through a
programmed dependence of their propulsion speed on the local density,
but also describes, to reasonable accuracy, the case where slowing
down is caused instead by collisions. We then considered the role of
gradient corrections, arising either from nonlocality in the
dependence of speed on density, or as additional terms in a combined
expansion in spatial gradients and angular harmonics of the local
distribution of particle orientations. Although the latter could in
principle cause qualitative shifts in phase behaviour, in common with
previous studies we do not find compelling evidence for anything more
than a quantitative shift numerically. Specifically, although the
``equilibrium" conditions for phase coexistence (those derivable from
the local part of the large deviation functional) are violated, the
system still exhibits conventional binodals whose defining property is
that the density of coexisting phases is independent of the
intermediate global density of the system. For the case where MIPS is
caused by collisional rather than programmed speed reduction, we find,
somewhat surprisingly, that there is once again almost no difference
in phase behaviour between ABPs and RTPs with matched angular
relaxation time. This is despite the fact that both the collisional
dynamics, and the presence of sharp interfaces in the system, involve
length scales below those at which the equivalence of the two models
can be formally established by systematic coarse graining.

\noindent{\em Acknowledgements:} The authors thank Rosalind Allen,
Ludovic Berthier, Cecile Cottin-Bizonne, Paul Goldbart, Davide
Marenduzzo, Joakim Stenhammar and Raphael Wittkowski for
discussions. MEC thanks the Royal Society for a Research
Professorship. This work was supported in part by EPSRC Grant
EP/J007404/1.

\appendix{}
\section{Moment-expansion}
\label{app:moments}

We start from the master equation on the probability density
$\psi(\r,\u,t)$:
\begin{equation}\label{app:EoM1}
  \dot\psi = - \n\cdot[v\u\psi]+D_r\Delta_\u \psi+\n\cdot [D_t \n \psi]-\alpha \psi +\frac{\alpha}\Omega \int \dO \psi(u')
\end{equation}
with $\Delta_\u$ the rotational part of the Laplacian acting on $\u$. We then use the
decomposition~\eqref{decomp} of $\psi$, which we recall here for
clarity
\begin{equation}\label{app:decomp}
    \psi(\r,\u,t) = \varphi + \p\cdot\u +{\bf Q}:(\u\u-{\bf I}/d)+ \Theta[\psi]
\end{equation}
In the following we will use the notation $M_{ab}=u_a
u_b -\delta_{ab}/d$ and the convention that repeated
indices are implictly summed upon: $f_i g_i\equiv \sum_i f_i g_i$. Because
$Q$ enters $\psi$ solely through the combination
$\sum_{i,j}Q_{ij}M_{ij}$, it can always be chosen traceless
symmetric. (Indeed, replacing $Q$ by $(Q+Q^\dagger)/2$ or $Q-\mathbf{1}
\Tr(Q)/d$ does not change this sum.)

We will use the standard notation for the scalar product on the sphere
\begin{equation}
  \la f,g \ra = \int \dO f(u) g(u)
\end{equation}
for which spherical (and angular) harmonics are orthogonal. To obtain
equations for $\dot \varphi$, $\dot \p$, $\dot {\bf Q}$, we take the
scalar product of equation~\eqref{app:EoM1} with $1$, $\bf u$ and
$M$. To do so, we compute
\begin{equation}
  \begin{aligned}
    \la 1, \psi \ra = \Omega \varphi;\quad
    \la \u, \psi \ra = \frac{\Omega}{d} \p;\quad
    \la  M_{a b}, \psi\ra = \frac{\Omega}d B_{abcd} Q_{cd}=\frac{\Omega}d \frac{2}{d+2} Q_{ab}
  \end{aligned}
\end{equation}
which rely on
\begin{equation}
  \begin{aligned}
    \la u_i,u_j \ra&=\frac{\Omega}d \delta_{ij};\qquad
  \la  u_i u_j, u_k u_\ell\ra = \frac{\Omega}{d (d+2)} (\delta_{ij}\delta_{k\ell}+\delta_{i\ell}\delta_{kj}+\delta_{ik}\delta_{j\ell})\\
  B_{ijk\ell}&=\frac{d}{\Omega}\la M_{ij},M_{kl}\ra=\frac{1}{d+2}(\delta_{ik}\delta_{j\ell}+\delta_{i\ell}\delta_{jk}-\frac{2}d \delta_{ij}\delta_{kl})\label{eqn:Bexplicit}
\end{aligned}
\end{equation}

Projecting \eqref{app:EoM1} on 1 then yields for $\dot \varphi$
\begin{equation}
   \hspace{1cm}\dot \varphi = -\frac{1}d \n [v\p]+\n\cdot [D_t \n \varphi]
\end{equation}
while projecting \eqref{app:EoM1} on $\u$ yields for $\dot \p$
\begin{equation}
\label{eqn:ugly1}
  \frac{\Omega \dot \p}{d}=-\la \u, \n [v \u \psi]\ra+ D_r \la  \u, \Delta_\u \psi \ra +\n\cdot [D_t \n \la  \u, \psi\ra] -\alpha \la \u, \psi\ra
\end{equation}
The last two terms are easy to compute since $\la \u, \psi\ra=\Omega
\p/d$. Then, spherical and angular harmonics of order $\ell$ are
eigenvectors of $\Delta_\u$ with eigenvalues $-\ell(\ell+1)$ and
$-\ell^2$, respectively, and the projection on $\u$ selects the $\p
. \u$ term so that
\begin{equation}
  D_r \la  \u, \Delta_u \psi\ra =-D_r (d-1) \Omega \frac{\p}d
\end{equation}
The first term of the r.h.s. of~\eqref{eqn:ugly1} is harder to compute since $u\psi$ is not
directly developed in harmonics. It can however be brought to a
simpler form:
\begin{equation}
  \begin{aligned}
    \la u_a, \n_b [v u_b \psi]\ra &=\n_b v  \la M_{ab}, \psi\ra + \n_b v \frac{\delta_{ab}}d \la 1, \psi\ra\\
    & = \frac{\Omega}d \n_b v  \frac{2}{d+2}Q_{ab}+ \n_a v \frac{ \Omega}{d} \varphi  \\
    & = \frac{2\Omega}{d(d+2)} \n_b v Q_{ab}+ \n_a v \frac{ \Omega}{d} \varphi  
  \end{aligned}
\end{equation}
Note that $\n$ always applies to everything on its right, unless
specified otherwise. One then obtain for the evoluation of $\p$
\begin{equation}\label{app:EoP}
   \dot p_a = -\n_a[ v \varphi] - \big(D_r(d-1)+\alpha\big) p_a +\n\cdot [D_t \n p_a] -\frac{2}{d+2} \nabla_b [v Q_{ab}]
\end{equation}

Next, one projects \eqref{app:EoM1} on $M_{ab}$ to obtain
\begin{equation}\label{app:EoQ}
  \la  M_{ab}, \dot \psi\ra = \frac{\Omega}d  \frac{2}{d+2}\dot Q_{ab}
\end{equation}
for the left hand side, and for the r.h.s.
\begin{equation}\label{eqn:rhsff}
  -\la  M_{ab}, \n v \u \psi\ra + D_r \la M_{ab}, \Delta_\u \psi\ra  + \n \cdot [D_t \n \la M_{ab}, \psi\ra] - \alpha \la M_{ab}, \psi \ra
\end{equation}
Again, the last two terms are simple since $\la M_{ab}, \psi\ra
=2\Omega Q_{ab}/[d(d+2)]$. Only the second angular/spherical harmonics
survive the projection on $M_{a b}$; they are eigenvectors of
$\Delta_\u$, with eigenvalues $-\ell(\ell+1)=2 d$ for d=3 and
$-\ell^2=-2d$ for $d=2$. Hence the rotational diffusion term
contributes a factor
\begin{equation}
  D_r \la  M_{ab}, \Delta_\u \psi\ra =-2 d D_r \frac{\Omega}d \frac 2 {d+2} Q_{ab}
\end{equation}
Again, the first term of~\eqref{eqn:rhsff} is slightly
more difficult to handle and has to be split as
\begin{equation}
  \begin{aligned}
    -\la M_{a b}, \n_c v u_c \psi \ra &= -\n_c v \la M_{a b},  u_c u_d p_d \ra -\n_c v \la M_{a b},  u_c \Theta[\psi] \ra \\
    &= -\n_c v p_d \la M_{a b}, (M_{cd}+\frac{d_{cd}}{d})\ra-\n_c v \frac{\Omega}d \tilde\chi_{abc}\\
    &= -\frac{\Omega}d \n_c v p_d B_{a bcd}-\n_c v \frac{\Omega}d \tilde\chi_{abc}
  \end{aligned}
\end{equation}
where we have introduced $\frac{\Omega}d \tilde\chi_{abc}=\la M_{a b},
u_c \Theta[\psi] \ra$ which comes from the projection of
$u\Theta[\psi]$ on the second harmonics and whose precise expression
we won't need.

Putting everything together then yields Eq.~\eqref{EoM4} of the main
text
\begin{equation}
\dot Q_{a b} =  - \frac{d+2}2\n_c v p_d B_{a bcd} -2 d D_r  Q_{ab} + \n\cdot[D_t \n Q_{ab}] - \alpha Q_{ab}-\n_c v\chi_{abc}
\end{equation}
with $\chi=(d+2)\tilde\chi/2$.

\section{From microscopic to mesoscopic It\=o-Langevin equations}
\paragraph{It\=o drift with functional dependences.}

\label{app:Itodrift}

The total gradient of $D$ with respect to $\r_\mu$, notated as
$D'(\r_\mu,[\rho])$ for lack of a better notation, is given by the
chain rule
\begin{equation}\label{app2:eq1}
  D'(\r_\mu,[\rho])\equiv \n_{1} D(\r_\mu,[\rho]) + \int d\r' \frac{\delta D(\r_\mu,[\rho])}{\delta\rho(\r')} \n_{\r_\mu} \rho(\r')
\end{equation}
where the first term on the r.h.s. comes from the explicit dependence
of $D$ on $\r_\mu$ and the second from the dependence of $D$ on
$\rho(\r')$ which itself depends on $\r_\mu$. By convention, $\n_{1}
D(\r_\mu,[\rho])$ is thus a ``partial gradient'' which acts upon the
explicit dependence of $D$ on $\r_\mu$ (its first argument) but not on
its implicit dependence through $\rho$. Using the explicit expression
of $\rho(\r')$, one has
\begin{equation}
  \label{app2:eq2}
  \n_{\r_\mu} \rho(\r')=\n_{\r_\mu} \sum_j \delta(\r'-\r_j)=\n_{\r_\mu}  \delta(\r'-\r_\mu)=-\n_{\r'} \delta(\r'-\r_\mu)
\end{equation}
Equation~\eqref{app2:eq2} shows why the notation $\n_{\r_\mu}$,
  which indicates that the gradient acts on $\r_\mu$ and not on $\r'$ or
  $\r_j$, is essential---though cumbersome---to get the correct result.
Inserting~\eqref{app2:eq2} into~\eqref{app2:eq1} and integrating by
parts, one then finds that
\begin{equation}
  \begin{aligned}
  \label{app2:eq3}
  D'(\r_\mu,[\rho])&=\n_{1}D(\r_\mu,[\rho]) + \int d\r'  \delta(\r'-\r_\mu) \n_{\r'}\frac{\delta D(\r_\mu,[\rho])}{\delta\rho(\r')}\\
  &=\n_{1}D(\r_\mu,[\rho]) +  \left(\n_{\r_\mu}\frac{\delta}{\delta\rho(\r_\mu)}\right) D(\r_\mu,[\rho])\\
\end{aligned}
\end{equation}

As a consistency check, and to understand how the operator
$\n_{\r_\mu}\frac{\delta}{\delta\rho(\r_\mu)}$ works, let us consider a
simple example where the diffusivity of a particle at position $\r_\mu$
is a linear functional of the density of its neighbours, convoluted by
some kernel $K$:
\begin{equation}
  D(\r_\mu,[\rho])=\int d\r' \rho(\r') K(\r_\mu-\r')=\sum_j K(\r_\mu-\r_j)
\end{equation}
In such a case, the total gradient of $D$ with respect to $\r_\mu$ can be
directly computed:
\begin{equation}\label{eqn:simplder}
  D'(\r_\mu,[\rho])=\sum_{j\neq \mu} \n K(\r_\mu-\r_j),
\end{equation}
where $\n K$ is the standard gradient of the function $K(\r)$. Applying
the formula~\eqref{app2:eq3}, one finds
\begin{equation}
  D'(\r_\mu,[\rho])=\int d\r' \rho(\r') \n K(\r_\mu-\r') + \int d\r' [\n_{\r_\mu} \frac{\delta \rho(\r')}{\delta \rho(\r_\mu)}] K(\r_\mu-\r')
\end{equation}
where the first term comes from the explicit derivative and the second
one from the functional derivative. Note that, in the latter
term, $\n_{\r_\mu}$ \textit{does not} act upon $K(\r_\mu-\r')$. The
computation can now be readily pursued, to give
\begin{equation}
  D'(\r_\mu,[\rho])=\int d\r' \sum_j \delta(\r'-\r_j) \n K(\r_\mu-\r') + \int d\r' [\n_{\r_\mu} \delta(\r'-\r_\mu)] K(\r_\mu-\r')\\
\end{equation}
Using again that $\n_{\r_\mu}\delta(\r'-\r_\mu)=-\n_{\r'}\delta(\r'-\r_\mu)$ and integrating by parts, one gets
\begin{equation}
  D'(\r_\mu,[\rho])= \sum_j \n K(\r_\mu-\r_j) -\n K(0) = \sum_{j\neq \mu} \n K(\r_\mu-\r_j)
\end{equation}

Interestingly, the functional derivative generates a term $-\n K(0)$
which will be absent whenever $K(r)$ is a symmetric kernel or when the
particles are \textit{not} self-interacting. (By this we mean that
the diffusivity $D$ of particle $\mu$ is a function of $\r_\mu$
and a functional of $\rho({\bf r})-\delta({\bf r}-\r_\mu)$; as such it
is functionally dependent only on the density field of other
particles, which differs from the total density by the
$\delta$-function self-term.) For complete generality however, we must
retain this term in the Langevin equation~\eqref{eqn:Langmicro} and
throughout the construction of the Langevin equation for $\rho(\r)$,
in order to derive a Fokker-Planck equation which is properly ordered
as this is crucial to get the correct steady-state. Since the
contribution of this term vanishes in many cases, it is often silently
omitted in the literature.

Let us further note that, looking at the simple derivation of
equation~\eqref{eqn:simplder}, all ambiguities can be overcome by
coming back to the microscopic definition of $\rho(\r)$. As often with
functionals, the notational problem is present only when working at
the field level, where it stems from an underlying ambiguity
concerning the gradient symbol $\n$, which is used in different
contexts to represent either a total derivative (acting on both
implicit and explicit dependences of $D(\r,[\rho])$, and hence
containing in practice a functional derivative) or a partial
derivative acting only on the first argument of $D$. This ambiguity
translates into an ordering problem at the level of the Fokker-Planck
equation. However, at the large deviation level---which corresponds
here to the large size limit---we are effectively interested in the
small noise limit of our stochastic partial differential
equations. The various ordering of the Fokker-Planck equation differ
by terms, similar to those that distinguish It\=o and Stratonovich
time-discretisation, which are generally
negligible~\cite{Tailleur2008JPA}. This is why the issue was for
instance (rightly) neglected in~\cite{Peruani2014JSM}.

\paragraph{Mesoscopic Langevin equation.}

Let us now consider a function $f(\r)$ and compute the time evolution of
$f(\r_\mu(t))$ where $\r_\mu(t)$ is solution of the Langevin
equation~\eqref{eqn:Langmicro}. Using It\=o's formula~\cite{Risken,oksendal}, one finds
\begin{equation}\label{eqn:Ito1}
  \dot f(\r_\mu(t))= \big[{\bf A} +\sqrt{2D}{\boldsymbol \eta}\big] \n f(\r_\mu) +D \Delta f(\r_\mu)
\end{equation}
where
\begin{equation}
{\bf A}=\V + \n_1 D(\r_\mu,[\rho])+\big(\n_{\r_\mu} \frac{\delta}{\delta \rho(\r_\mu)}\big) D(\r_\mu,[\rho])
\end{equation}
Introducing $\rho_\mu=\delta(\r-\r_\mu)$ and using that, for any
function $H$, $H(\r_\mu)=\int \d\r \rho_\mu H(\r)$,
Eq.~\eqref{eqn:Ito1} can be rewritten
\begin{equation}\label{eqn:astuce}
  \dot f(\r_\mu(t))= \int \d\r \rho_\mu(\r,t)[(\A+\sqrt{2D}{\boldsymbol\eta}) \grad f(\r) +D(\r,[\rho]) \Delta f(\r)]
\end{equation}
where the derivatives in $\grad f(\r)$ and $\Delta f(\r)$ are now
taken with respect to $\r$. The last term in \eqref{eqn:astuce},
  even though it looks harmless, requires some explanations. Indeed,
  $D(\r_\mu,[\rho])$ has dependencies on $\r_\mu$ both through its
  explicit dependence and through its functional dependence on
  $\rho(\r')=\sum_\alpha \delta(\r'-\r_\alpha)$. When going
  from~\eqref{eqn:Ito1} to~\eqref{eqn:astuce}, we use that
\begin{equation}\label{eqn:bla}
  \delta(\r-\r_\mu) D(\r_\mu,[\rho]) \Delta f(\r_\mu) =  \delta(\r-\r_\mu) D(\r,[\rho]) \Delta f(\r),
\end{equation}
where we have replaced all the $\r_\mu$'s by $\r$ \textit{but the one
  in $\rho$}, i.e., $\rho(\r')=\sum_\alpha \delta(\r'-\r_\alpha)$ has
not been replaced by $\delta(\r'-\r)+\sum_{\alpha\neq \mu}
\delta(\r'-\r_\alpha)$. There is thus no dependence of $D$ on $\r$
through $\rho$. From now on, gradients $\n D(\r,[\rho])$ are not
ambiguous anymore: they solely apply to the first argument of $D$
which is the only place where $D$ depends on $\r$.  There is thus no
need anymore for the notation $\grad_1$.  Integrating by part
Eq.~\eqref{eqn:astuce} then leads to
\begin{equation}
  \label{eqn:fdot1}
  \dot f(\r_\mu(t))= \int \d\r f(\r) \n\cdot [- (\A+\sqrt{2D}{\boldsymbol\eta})\rho_\mu(\r,t) + \n (D\rho_\mu)]
\end{equation}
Alternatively, $\dot f(\r_\mu)$ can also be written
\begin{equation}\label{eqn:fdot2}
  \dot f(\r_\mu)= \int \d \r   \dot \rho_\mu(\r) f(\r)
\end{equation}
Since the equations~\eqref{eqn:fdot1} and~\eqref{eqn:fdot2} hold for
any function $f$, one gets
\begin{equation}
  \dot \rho_\mu=\n [- (\A+\sqrt{2D}{\boldsymbol\eta})\rho_\mu(\r,t) + \n (D\rho_\mu)]
\end{equation}
Introducing the Gaussian white noise 
\begin{equation}
  \sqrt{2 D \rho}\Lambda\equiv\sum_\mu \sqrt{2D} {\boldsymbol \eta_\mu}\rho_\mu
\end{equation}
whose statistics satisfy
\begin{equation}
\la \Lambda \ra =0;\qquad \la \Lambda(\r,t)\Lambda(\r',t')\ra =\delta(t-t') \delta(\r-\r')
\end{equation}
one gets
\begin{equation}
  \dot \rho(\r) = \n [-\rho(\r) \V + D \n \rho(\r) -\rho \big(\n_{\r} \frac{\delta}{\delta \rho(\r)}\big) D - \sqrt{2 D \rho} \Lambda]
\end{equation}
which is Eq.~\eqref{coll} as required.

\section{Functional Fokker-Planck Equation}

\label{app:FP}

The easiest way to derive the Fokker-Planck equation~\eqref{eq:FPErho}
starting from the Langevin equation~\eqref{coll} is to spatially
discretize the latter in one spatial dimension
\begin{equation}
  \dot \rho_i = A_i +\frac 1 2 (B_{i,i+1}\eta_{i+1} + B_{i,i-1}\eta_{i-1})
\end{equation}
where
\begin{align}
  A_i&=\n_i\big[-\rho_{i}V_{i}+\frac 1 2 D_i (\rho_{i+1}-\rho_{i-1})+\frac 1 2 \big(\frac{\partial D_i}{\partial\rho_{i+1}}-\frac{\partial D_i}{\partial\rho_{i-1}}\big)\rho_i\big]\label{app:defA}\\
  B_{i,i+1}&=-\frac 1 2 \sqrt{2 D_{i+1}\rho_{i+1}}\qquad\text{and}\qquad B_{i,i-1}=\frac 1 2 \sqrt{2 D_{i-1}\rho_{i-1}}
\end{align}
Note that we use centred differences $\n_i O_i\equiv \frac 1 2
(O_{i+1}-O_{i-1})$ to respect the isotropy of the equation. One can then use the
general relation between It\=o-Langevin dynamics and the Fokker-Planck
equation~\cite{Risken},
\begin{equation}
  \dot x_i=A_i+B_{ij}\eta_j\quad\longrightarrow\quad \dot P(x)=-\sum_i\frac{\partial}{\partial x_i}A_iP +\frac 1 2 \sum_{i,j,k} \frac{\partial}{\partial x_i} \frac{\partial}{\partial x_j} B_{ik}B_{jk}P
\end{equation}
to derive the Fokker-Planck equation satisfied by $P({\rho_i})$. To do
so, we first note that
\begin{align}
  2 \sum_k B_{ik} B_{jk} &= \sum_k \left[\sqrt{D_{i+1}\rho_{i+1}} \delta_{k,i+1}+ \sqrt{D_{i-1}\rho_{i-1}}\delta_{k,i-1}\right]\nonumber\\&\qquad \times \: \left[\sqrt{D_{j+1}\rho_{j+1}} \delta_{k,j+1}+ \sqrt{D_{j-1}\rho_{j-1}}\delta_{k,j-1}\right]\\
  &=\sqrt{D_{i+1}\rho_{i+1}D_{j+1}\rho_{j+1}}\delta_{i,j}-\sqrt{D_{i-1}\rho_{i-1} D_{j+1}\rho_{j+1}}\delta_{i-2,j}\nonumber\\&\:\quad-\sqrt{D_{i+1}\rho_{i+1}D_{j-1}\rho_{j-1}}\delta_{i+2,j}+\sqrt{D_{i-1}\rho_{i-1}2 D_{j-1}\rho_{j-1}}\delta_{i,j}
\end{align}
To (slightly) lighten the notation, we write $\partial_{\rho_i}$ for the
operator $\frac{\partial}{\partial \rho_i}$. The second order
differential operator then becomes
\begin{align}
  2\sum_{i,j,k} \partial_{\rho_i} \partial_{\rho_j} B_{ik}B_{jk}&= \sum_i \partial_{\rho_i}\big[\partial_{\rho_i} D_{i+1}\rho_{i+1}-\partial_{\rho_{i-2}} D_{i-1}\rho_{i-1}\nonumber\\&\qquad\qquad\qquad-\partial_{\rho_{i+2}} D_{i+1}\rho_{i+1}+\partial_{\rho_i} D_{i-1}\rho_{i-1} \big]\nonumber\\
   &=\sum_i \partial_{\rho_i}\big(\partial_{\rho_i}-\partial_{\rho_{i+2}}\big) D_{i+1}\rho_{i+1}-\sum_i\partial_{\rho_{i}} \big(\partial_{\rho_{i-2}}-\partial_{\rho_{i}}\big) D_{i-1}\rho_{i-1}\nonumber
 \end{align}
Shifting by $\pm1$ the indices in the two sums, one finally gets
\begin{equation}
  2\sum_{i,j,k} \partial_{\rho_i} \partial_{\rho_j} B_{ik}B_{jk}= \sum_i \left(\frac{\partial}{\partial \rho_{i-1}}-\frac{\partial}{\partial \rho_{i+1}}\right) \left(\frac{\partial}{\partial \rho_{i-1}}-\frac{\partial}{\partial \rho_{i+1}}\right) D_i \rho_i
\end{equation}

We now have to compute the drift term in the Fokker-Planck equation
$\sum_i \partial_{\rho_i} A_i$. Noting that $A_i$ is a gradient,
$A_i=\n_i C_i$, where $C_i$ can be read in~\eqref{app:defA}, the drift
term can be rewritten
\begin{align}
  2\sum_i \partial_{\rho_i} A_i&=   \sum_i \partial_{\rho_i} (C_{i+1}-C_{i-1})=\sum_i (\partial_{\rho_{i-1}}-\partial_{\rho_{i+1}})C_{i}\\
  &=\sum_i (\partial_{\rho_{i-1}}-\partial_{\rho_{i+1}})\left(-\rho_i V_i +D_i\frac{\rho_{i+1}-\rho_{i-1}}2 +\frac 1 2 \left(\frac{\partial D_i}{\partial \rho_{i+1}}-\frac{\partial D_i}{\partial \rho_{i-1}}\right)\rho_i\right)\nonumber
\end{align}
Putting everything together yields
\begin{equation}\label{app:FPinter}
  \dot P = \sum_i \frac {\partial_{\rho_{i-1}}-\partial_{\rho_{i+1}}}2 \left(\rho_i V_i -D_i\n_i\rho_{i} -\frac {\rho_i}2 \Big(\frac{\partial D_i}{\partial \rho_{i+1}}-\frac{\partial D_i}{\partial \rho_{i-1}}\Big)+\frac{\partial_{\rho_{i-1}}-\partial_{\rho_{i+1}}}2 D_i\rho_i \right)
\end{equation}
Let us now note that $\rho_i$ commutes with
$\partial_{\rho_{i-1}}-\partial_{\rho_{i+1}}$ but not $D_i$, since the latter
is in principle a function of all the $\rho_j$'s. However, the
application of $\partial_{\rho_{i-1}}-\partial_{\rho_{i+1}}$ on $D_i$
exactly cancels the one before last term in~\eqref{app:FPinter}, so that
\begin{equation*}
  \dot P = \sum_i \frac 1 2 (\partial_{\rho_{i-1}}-\partial_{\rho_{i+1}}) \left(\rho_i V_i -D_i\n_i\rho_{i} + D_i \rho_i \frac{\partial_{\rho_{i-1}}-\partial_{\rho_{i+1}}}2 \right)P
\end{equation*}
Taking at this stage the continuum limit and integrating once by part 
 yields the correct Fokker-Planck equation
\begin{equation}\label{app:FP1d}
  \dot P = - \int \d x  \frac{\partial}{\partial \rho(x)} \n \left(\rho(x) V(x,[\rho]) -D(x,[\rho]) \n \rho - D(x,[\rho]) \rho \n_x \frac{\delta}{\delta \rho(x)} \right)P
\end{equation}
where one has recognised the operator
\begin{equation}\label{app:deepmeaning}
  \frac 1 2 (\partial_{\rho_{i-1}}-\partial_{\rho_{i+1}})= \n_i \frac{\partial}{\partial \rho_i} \longrightarrow \n_x \frac{\delta}{\delta \rho(x)}
\end{equation}
Equation~\eqref{app:deepmeaning} is actually the best way to make
sense of this operator and to see that the gradient really applies to the
field with respect to which we are taking a functional derivative and
not to the argument of this functional derivative.  The
equation~\eqref{app:FP1d} directly generalizes to~\eqref{eq:FPErho} in
higher dimensions.  We note that to get the correct ordering between
$\rho D$ and $\n_\r\frac{\delta}{\delta \rho(\r)}$ in the Fokker-Planck
equation, it was necessary to correctly account for the atypical form of the
spurious It\=o drift as defined in the main text.

\end{document}